\documentclass[fleqn,usenatbib]{mnras}

\usepackage{newtxtext,newtxmath}
\usepackage[T1]{fontenc}

\DeclareRobustCommand{\VAN}[3]{#2}
\let\VANthebibliography\thebibliography
\def\thebibliography{\DeclareRobustCommand{\VAN}[3]{##3}\VANthebibliography}

\usepackage{graphicx}	% Including figure files
\usepackage{amsmath}	% Advanced maths commands
\usepackage{comment}
\title[Meteorites from near-Earth asteroids]{Identifying parent bodies of meteorites among near-Earth asteroids}

\author[A. Carbognani and M. Fenucci]{
A. Carbognani$^{1}$\thanks{E-mail: - }
and M. Fenucci$^{2,3}$
\\
$^{1}$INAF - Osservatorio di Astrofisica e Scienza dello Spazio, Via Gobetti 93/3, 40129 Bologna, Italy\\
$^{2}$ESA ESRIN / PDO / NEO Coordination Centre, Largo Galileo Galilei, 1, 00044 Frascati (RM), Italy\\
$^{3}$Elecnor Deimos, Via Giuseppe Verdi, 6, 28060 San Pietro Mosezzo (NO), Italy
}

\date{Accepted XXX. Received YYY; in original form ZZZ}

\pubyear{2015}

\begin{document}
\label{firstpage}
\pagerange{\pageref{firstpage}--\pageref{lastpage}}
\maketitle

\begin{abstract}
Meteorites provide an important source of information about the formation and composition of asteroids, because the level of accuracy of studies and analyses performed in a laboratory cannot be achieved by any ground or space based observation. To better understand what asteroid types a meteorite represents, it is crucial to identify the body they originated from. 
In this paper, we aim to determine possible parent bodies for the known meteorite falls among the known population of near-Earth asteroids (NEAs). By using the similarity criterion $D_N$, based on geocentric quantities, we found 20 possible NEA-meteorite pairs. By performing additional numerical simulations of the backward dynamics, we found that 12 of these pairs may be associated with a possible separation event from the progenitor NEA, while the remaining 8 pairs appear to be ambiguous or random associations. 
The most interesting are the Pribram and Neuschwanstein meteorites, which are dynamically associated with (482488) 2012 SW20 with a common separation age dating back to about 20$-$30 kyr ago, and the Motopi Pan meteorite, that has three candidate parent bodies: (454100) 2013 BO73, 2017 MC3, and 2009 FZ4. The average time of separation between our meteorite list and the progenitor body appears to be about 10 kyr, a time consistent with what is expected from the collision frequency of small NEAs. Based on our results, we suggest that about 25 per cent of meteorites do not originate in the main belt, but mainly from little collision events happening between NEAs in the inner Solar System. 
\end{abstract}

\begin{keywords}
meteorites, meteors, meteoroids -- minor planets, asteroids: general -- celestial mechanics 
\end{keywords}

%%%%%%%%%%%%%%%%%%%%%%%%%%%%%%%%%%%%%%%%%%%%%%%%%%

%%%%%%%%%%%%%%%%% BODY OF PAPER %%%%%%%%%%%%%%%%%%

\section{Introduction}
\begin{figure}
\centering
\includegraphics[trim={3.5cm 0 2.7cm 0},clip, width=0.5\textwidth]{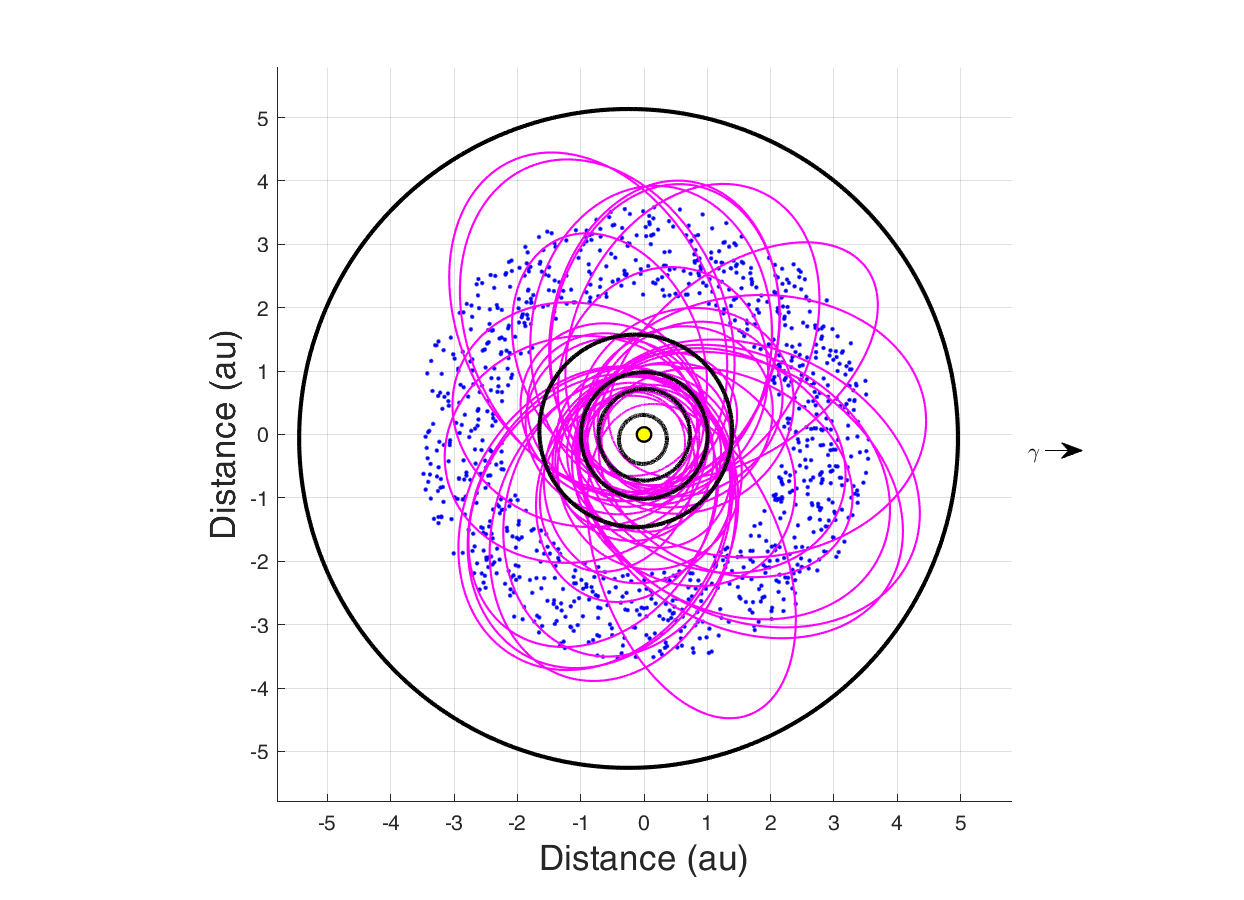}
\caption{The meteorites orbits of Table~\ref{tab:meteorites} projected on the ecliptic plane. The outer circle is the Jupiter's orbit. The main belt is represented by dots with a random distribution.}
\label{fig:meteorites_orbits}
\end{figure}

The Earth's atmosphere is continuously bombarded by small near-Earth-type asteroids. According to NASA's CNEOS data center\footnote{\url{https://cneos.jpl.nasa.gov/}}, there is an event detectable by military satellites on average every two weeks, with a mean diameter of about 1-2 m. About 9.8\% of the events belong to Jupiter Family Comets, while 85.5\% have a Tisserand parameter with respect to Jupiter typical of asteroid orbits \citep{Pena-Asensio2022}. \\
These meteoroids enter the atmosphere at speeds of several tens of km per second, give rise to brilliant fireballs, and disintegrate as they fall into the deep atmosphere due to the compression exerted on meteoroid by the shock wave. The first disintegration occurs at variable altitudes according to the strength of the body (0.1$-$4 MPa), while the extinction occurs at altitudes of some tens of km \citep{Borovicka2008}. For fireballs arriving at lower altitudes, between 20 and 30 km in height, there is a good chance that the fragments of the original meteoroid begin the dark flight phase and give rise to the recovery of meteorites on the ground.\\
From the heliocentric orbits of the meteoroids associated with meteorites, we know that they originate from the asteroids population, but it is difficult to identify which are the progenitors. Notable exceptions are the asteroid Vesta and meteorites of the HED (Howardite$-$Eucrite$-$Diogenite) type. In this case the association was facilitated by the optical reflectance spectral features similar to those of Vesta and other smaller asteroids belonging to the Vesta family \citep{Binzel1993}. In general, the problem is not so easy. For example, the progenitor bodies of the ordinary chondrites (OC) which make up about 80\% of the falls are S-type asteroids, but they are not yet identified with certainty. OCs must have originated from at least three progenitor bodies, so as to explain the presence of chondrites of type H (42.5\% of the OCs), L (46.2\%) and LL (11.3\%) according to the decreasing content of iron and metals \citep{Vernazza2015}.\\
According to the current paradigm, the progenitor meteoroids of meteorites were formed millions of years ago following collisions between main-belt asteroids that created asteroid families. Subsequently, due to the mean-motion orbital resonances with Jupiter and the Yarkovsky effect, they were placed on near-Earth type orbits which led them to fall on Earth \citep[see e.g.][and references therein]{Bottke2002, Granvik2017, Granvik2018}. For example, this would be the origin of some L chondrites (Creston, Novato, Innisfree, and Jesenice) and the rare EL enstatite chondrite associated to main belt asteroid (161) Athor \citep{Jenniskens2019, Avdellidou2022}. The prominent Flora family in the inner part of the main belt is a good candidate to be the origin of rare LL-type ordinary chondrites \citep{Vernazza2008}, while the large S-type asteroid (6) Hebe, located adjacent to both the $\nu_6$ secular resonance with Saturn and 3:1 mean-motion resonance with Jupiter is the probable source of H-type OC \citep{Morbidelli1994, Gaffey1998}. Also \cite{Granvik2018b}, recalculating the orbits of 25 known meteorite falls based on the trajectory information reported in the literature, finds that most meteorite originate in the inner main asteroid belt and escape through the 3:1 mean-motion resonance or the $\nu_6$ secular resonance.\\
However, it cannot be excluded that a part of the meteorites also originates directly from the dynamic population of the near-Earth asteroids (NEAs). Possible formation mechanisms include collisions with smaller NEAs, thermal fragmentation \citep{Delbo2014, Granvik2016}, partial rotational disaggregation of rubble piles due to YORP effect \citep{Scheeres2018}, tidal disruption of rubble piles during flyby with planets \citep{Zhang2020}, sublimation of small reserves of volatile materials, or as a result of activity mechanisms not yet well understood on the asteroid's surface as for (3200) Phaethon, the progenitor of the Geminid meteoroid stream \citep{Whipple1983, Williams1993}, and (101955) Bennu \citep{Melikyan2021}. In the case of this last NEA, the activity of ejecting stones into space is detectable only from spacecraft, but not through observations from the ground because the intensity is too low.\\
The rotational disaggregation due to YORP effect appears to be the primary formation processes for small close binaries between the population of NEAs: the estimated proportion of binaries among NEAs larger than 0.3 km is $15\pm 4$\% \citep{Pravec2006}. Pairs of asteroids share similar heliocentric orbits but they are not bound together: integrations of their orbits indicated that they separated with low relative velocities, so they can be considered as unborn binaries with the same physical formation processes \citep{Pravec2010}. Most of the known pairs are found between main-belt asteroids \citep{Kyrylenko2021}, but at least two pairs of asteroids (2015 EE7$-$2015 FP124 and 2017 SN16$-$2018 RY7), have also been identified in the NEAs population, suggesting a common origin and a very recent separation between the two components, estimated in about $10$ kyr for the second pair \citep{Moskovitz2019}. In both cases $-$ binary or pairs $-$ it is possible to have the dispersion, on the original orbit, of large meteoroids capable of generating fireballs and meteorites. All the physical processes mentioned above fall in the heterogeneous category of active asteroids, transversal to both main belts and NEAs \citep{Jewitt2012}: for a recent review about meteor showers from active asteroids and dormant comets between NEAs see \citet{Quanzhi2018}. \\
One objection to this hypothesis of connection between NEAs and meteorites may be that NEAs cannot be progenitors of meteoroids producing meteorites because about 2/3 of the NEAs have spectra compatible with ordinary LL-type chondrites which are a minority compared to other types of meteorites \citep{Vernazza2015}. While this may be true in a statistical sense, it does not rule out the possibility that there are some NEAs that, by means of some of the physical processes mentioned above, can release large meteoroids into interplanetary space forming faint meteoroids stream along the orbit. Furthermore, considering that the NEA population contains asteroids that pose a risk to the Earth and that the mitigation strategy depends on the composition of the object, try associating meteorites and NEAs can be a good way to get low-cost information on potentially dangerous asteroids and the meteorite-NEAs association may become very important also in view of the mining of asteroids because it is possible to have useful information to decide which asteroids to exploit \citep{OLeary1977}. Recently, a dynamical association between the impactor 2018LA and the Potentially Hazardous Asteroid (454100) 2013 BO73 have been proposed \citep{deLaFuente2019} and, as we see later, we find the same association. \\
So the question we want to answer is: are there NEAs that could be the progenitors - or brothers - of some of the meteorites known today? To answer this question, we investigated the possible orbital connection between meteorites and NEAs in order to identify those that are the most probable progenitors of meteorites whose heliocentric orbit is known. For this reason, only the meteorites whose fireball was directly triangulated during the atmospheric fall were taken into consideration, see Table~\ref{tab:meteorites} for a list and Figure~\ref{fig:meteorites_orbits} for heliocentric orbits plot. \\

The paper is organized as follow: in Section~\ref{sec:orbit} we will see the adopted criterion for orbits similarity. In Section~\ref{sec:association} we will see the list of the most probable associations between meteorites and NEAs, while in Section~\ref{sec:integration} we will see the results about the numerical integration of the orbits to dynamically establish the connection between NEAs and meteorites. In Section~\ref{s:discussion} we discuss the results obtained and finally we provide our conclusions.

\begin{table*}
	\centering
	\caption{List of meteorites with a heliocentric orbit from fireballs triangulation sorted by date of fall in Julian Days (JD): $a$ = semi major axis; $e$ = eccentricity; $i$ = orbit inclination; $\Omega$ = longitude of ascending node; $\omega$ = argument of perihelion; $M$ = mean anomaly of the fall. Orbital elements for equinox J2000.0. Meteorite type: OC = ordinary chondrite; C = carbonaceous chondrite; EC= enstatite chondrite; AE= achondrite Eucrite; AH=achondrite Howardite; U=Ureilite. In this table meteorites type are from  Meteoritical Bulletin Database (\url{https://www.lpi.usra.edu/meteor/}). For an extensive set of references regarding this list of meteorites you can consult the Meteorite Orbits.info website (\url{https://www.meteoriteorbits.info/}). These databases were last consulted on May 10, 2023.} 
	\label{tab:meteorites}
        \setlength\tabcolsep{3pt} % default value: 6pt
	\begin{tabular}{llcccccccc} 
		\hline
		N & Name & Type & $a$ (au) & $e$ & $i$ ($^\circ$) & $\Omega$ ($^\circ$) & $\omega$ ($^\circ$) & $M$ ($^\circ$) & JD fall\\
		\hline
01  &  Pribram                  & OC H5    & 2.40$\pm$0.002  &  0.6711$\pm$0.0003 &  10.482$\pm$0.0004  &  17.79147$\pm$0.00001  & 241.75$\pm$0.013    & 349.539 &  2436666.333\\
02  &  Lost City                & OC H5    & 1.66$\pm$0.05   &   0.417$\pm$0.005  &  12.0$\pm$0.5       &  283.0$\pm$0.5         & 161.0$\pm$0.5       & 7.182   &  2440591.333\\
03  &  Innisfree                & OC L5    & 1.87$\pm$0.005  &  0.4732$\pm$0.0005 &  12.27$\pm$0.05     &  316.80$\pm$0.05       & 177.97$\pm$0.05     & 0.640   &  2443181.292\\   
04  &  Benesov                  & OC LL3.5 & 2.48$\pm$0.002  &  0.6274$\pm$0.0004 &  23.981$\pm$0.007   &  47.0009$\pm$0.0001    & 218.370$\pm$0.008   & 352.417 &  2448384.458\\
05  &  Peekskill                & OC H6    & 1.49$\pm$0.03   &  0.41$\pm$0.01     &  4.9$\pm$0.2        &  17.030$\pm$0.001      & 308.0$\pm$1.0       &  21.540 &  2448905.292\\
06  &  Tagish Lake              & C2-ung   & 2.1$\pm$0.2     &  0.57$\pm$0.05     &  1.4$\pm$0.9        &  297.900$\pm$0.003     & 222.0$\pm$2.0       & 349.895 &  2451561.833\\
07  &  Moravka                  & OC H5    & 1.85$\pm$0.07   &  0.47$\pm$0.02     &  32.2$\pm$0.5       &  46.2580$\pm$0.00005   & 203.5$\pm$0.6       & 352.385 &  2451671.042\\  
08  &  Neuschwanstein           & EC EL6   & 2.40$\pm$0.02   &  0.670$\pm$0.002   &  11.41$\pm$0.03     &   16.82664$\pm$0.00001 & 241.20$\pm$0.06     & 349.420 &  2452371.347\\
09  &  Park Forest              & OC L5    & 2.53$\pm$0.19   &  0.680$\pm$0.023   &  3.2$\pm$0.3        &   6.1156$\pm$0.0007    & 237.5$\pm$1.6       & 350.715 &  2452725.493\\  
10  &  Villalbeto de la Peña    & OC L6    & 2.3$\pm$0.2     &  0.63$\pm$0.04     &  0.0$\pm$0.2        &  283.6712$\pm$0.00005  & 132.3$\pm$1.5       &   9.229 &  2453009.199\\  
11  &  Bunburra Rockhole        & AE       & 0.85$\pm$0.0004 &  0.2427$\pm$0.0005 &   8.95$\pm$0.03     &  297.595$\pm$0.0005    & 210.04$\pm$0.06     & 133.571 &  2454302.301\\
12  &  Almahata Sitta           & U        & 1.308201$\pm$0  &  0.31206$\pm$0     &   2.54220$\pm$0     &  194.101138$\pm$0      & 234.44897$\pm$0     & 330.834 &  2454746.615\\
13  &  Buzzard Coulee           & OC H4    & 1.25$\pm$0.02   &  0.23$\pm$0.02     &  25.0$\pm$0.8       &  238.93739$\pm$0.00008 & 211.3$\pm$1.4       & 340.573 &  2454791.518\\
14  &  Maribo                   & CM2      & 2.43$\pm$ 0.2   &  0.805$\pm$0.011   &   0.25$\pm$0.16     &  297.46$\pm$0.15       & 279.4$\pm$0.6       & 348.713 &  2454849.299\\
15  &  Jesenice                 & OC L6    & 1.75$\pm$0.07   &  0.431$\pm$0.022   &  9.6$\pm$0.5        &  019.196$\pm$0.0005    & 190.5$\pm$0.5       & 356.220 &  2454930.583\\
16  &  Grimsby                  & OC H5    & 2.04$\pm$ 0.05  &  0.518$\pm$0.011   &  28.07$\pm$0.28     &  182.9561$\pm$0.00005  & 159.865$\pm$0.43    &   5.548 &  2455100.544\\
17  &  Ko\v sice                & OC H5    & 2.71$\pm$0.24   &  0.647$\pm$0.032   &  2.0$\pm$0.8        &  340.072$\pm$0.004     & 204.2$\pm$1.2       & 355.951 &  2455256.434\\
18  &  Mason Gully              & OC H5    & 2.470$\pm$0.004 &  0.6023$\pm$0.009  &  0.832$\pm$0.013    & 203.2112$\pm$0.00005   & 218.95$\pm$0.03     & 351.810 &  2455299.941\\
19  &  Kri\v zevci              & OC H6    & 1.544$\pm$0.01  &  0.521$\pm$0.004   &  0.64$\pm$0.03      & 315.55$\pm$0.01        & 254.4$\pm$0.1       & 335.381 &  2455597.472\\
20  &  Sutter's Mill            & C        & 2.59$\pm$0.35   &  0.824$\pm$0.02    &  2.38$\pm$1.16      &  32.77$\pm$0.06        &  77.8$\pm$3.2       &  10.452 &  2456040.118\\
21  &  Novato                   & OC L6    & 2.088$\pm$0.077 &  0.526$\pm$0.017   &  5.508$\pm$0.040    &  24.99$\pm$0.0035      & 347.352$\pm$0.134   &   3.360 &  2456218.613\\
22  &  Chelyabinsk              & OC LL5   & 1.72$\pm$0.02   &  0.571$\pm$0.006   &  4.98$\pm$0.12      & 326.459$\pm$0.001      & 107.67$\pm$0.17     &  20.000 &  2456338.640\\
23  &  Annama                   & OC H5    & 1.99$\pm$0.12   &  0.69 $\pm$0.02    & 14.65$\pm$0.46      &  28.611$\pm$0.001      & 264.77$\pm$0.55     & 344.111 &  2456766.426\\
24  &  \v Zd'ár nad Sázavou     & OC L3    & 2.093$\pm$0.006 &  0.6792$\pm$0.001  &  2.796$\pm$0.009    & 257.262$\pm$0.010      & 257.721$\pm$0.014   & 345.583 &  2457001.178\\
25  &  Porangaba                & OC L4    & 2.45$\pm$1.10   &  0.64$\pm$0.11     &  8.6$\pm$3.2        & 288.921$\pm$0.001      & 142.8$\pm$6.7       &   6.638 &  2457032.232\\
26  &  Sariçiçek                & AH       & 1.454$\pm$0.083 &  0.304$\pm$0.039   & 22.6$\pm$1.6        & 159.849$\pm$0.004      & 182.8$\pm$1.6       & 358.576 &  2457268.342\\
27  &  Creston                  & OC L6    & 1.300$\pm$0.019 &  0.410$\pm$0.013   &  4.228$\pm$0.070    &  30.458$\pm$0.006      &   7.20$\pm$0.13     & 323.239 &  2457319.741\\
28  &  Murrili                  & OC H5    & 2.521$\pm$0.075 &  0.609$\pm$0.012   &  3.32$\pm$0.060     &  64.742$\pm$0.0033     & 354.557$\pm$0.039   &   1.050 &  2457353.947\\
29  &  Ejby                     & OC H5/6  & 2.81$\pm$0.09   &  0.65$\pm$0.011    &  0.96$\pm$0.10      & 317.211$\pm$0.0001     & 197.75$\pm$0.10     & 357.102 &  2457425.380\\
30  &  Stubenberg               & OC LL6   & 1.525$\pm$0.010 &  0.395$\pm$0.004   &  2.07$\pm$0.03      & 346.520$\pm$0.0001     & 221.02$\pm$0.03     & 342.834 &  2457454.400\\
31  &  Dishchii'bikoh           & OC LL7   & 1.129$\pm$0.008 &  0.205$\pm$0.004   & 21.24$\pm$0.27      & 72.1206$\pm$0.0002     & 108.7$\pm$1.5       &  50.231 &  2457541.955\\
32  &  Dingle Dell              & OC LL6   & 2.254$\pm$0.034 &  0.5905$\pm$0.0063 &  4.051$\pm$0.012    & 218.252$\pm$0.00032    & 215.773$\pm$0.049   & 352.191 &  2457693.002\\
33  &  Hamburg                  & OC H4    & 2.73$\pm$0.05   &  0.661$\pm$0.006   &  0.604$\pm$0.11     & 296.421$\pm$0.03       & 211.65$\pm$0.3      & 354.949 &  2458135.547\\
34  &  Motopi Pan               & AH       & 1.3764$\pm$0.0001 & 0.43186$\pm$0.00006 & 4.2974$\pm$0.0004 & 71.869605$\pm$0.000012 & 256.04869$\pm$0.00055 & 327.170 &  2458272.197\\ 
35  &  Ozerki                   & OC L6    & 0.84$\pm$0.02   &  0.199$\pm$0.03    & 18.443$\pm$3.047    & 89.656$\pm$ ?          & 335.286$\pm$5.147   & 215.712 &  2458290.553\\ 
36  &  Viñales                  & OC L6    & 1.217$\pm$0.005 &  0.391$\pm$0.005   & 11.47$\pm$0.05      & 132.28 $\pm$0.005      & 276.97$\pm$0.05     & 306.489 &  2458516.262\\
37  &  Arpu Kuilpu              & OC H5    & 2.75$\pm$0.03   &  0.671$\pm$0.003   & 2.03$\pm$0.01       & 250.36$\pm$0.01        & 43.25$\pm$0.02      & 353.166 &  2458635.912\\ 
38  &  Flensburg                & C1       & 2.82$\pm$0.03   &  0.701$\pm$0.003   &  6.82$\pm$0.06      & 349.207$\pm$0.001      & 307.25$\pm$0.16     &   7.480 &  2458739.035\\ 
39  &  Cavezzo                  & OC L5-an & 1.82$\pm$0.22   &  0.46$\pm$0.063    &  4.0$\pm$1.6        & 280.52311$\pm$0.00001  & 179.2$\pm$4.8       &   0.261 &  2458850.268\\
40  &  Novo Mesto               & OC L5    & 1.451$\pm$0.004 &  0.6086$\pm$0.0006 &  8.755$\pm$0.063    & 338.993041$\pm$0.00001 &  82.649$\pm$0.184   &  28.826 &  2458907.896\\
41  &  Madura Cave              & OC L5    & 0.889$\pm$0.003 &  0.327$\pm$0.009   &  0.12$\pm$0.08      &  88.703764$\pm$0.00001 & 312.02$\pm$0.51     & 260.842 &  2459020.336\\
42  &  Traspena                 & OC L5    & 1.125$\pm$0.016 &  0.386$\pm$0.013   &  4.55$\pm$0.19      & 297.8270$\pm$0.0003    & 273.93$\pm$0.98     & 309.939 &  2459232.514\\
43  &  Winchcombe               & CM2      & 2.5855$\pm$0.0077 & 0.6183$\pm$0.0011 & 0.460$\pm$0.014     & 160.1955$\pm$0.0014    & 351.798$\pm$0.018  &   1.524 &  2459274.410\\
44  &  Antonin                  & OC L5    & 1.1269$\pm$0.0007 & 0.2285$\pm$0.0006 & 24.22$\pm$0.05      & 112.5807$\pm$0.0001    & 257.16$\pm$0.09    & 307.227 &  2459410.625\\
		\hline
	\end{tabular}
\end{table*}

\section{Adopted criterion for orbits similarity}
\label{sec:orbit}
In general, the \citet{Southworth1963} Dissimilarity Criterion $D$ or its variants \citep[see e.g.][]{Drumond1981, Jopek1993, Jenniskens2008, Rozek2011}, is used to search for a connection between the observed orbits of the meteors and the progenitor body. Usually a value $D \leq 0.1$ indicates a higher degree of similarity between two orbits. A first problem with this approach is that the meteor's heliocentric orbital elements are derived quantities and cannot always be determined with precision comparable to that of asteroids. A second problem is that orbits change over time due to planetary perturbations, so the value of $D$ may not be indicative of a common origin \citep{Quanzhi2018}. In the case of meteors and fireballs it is therefore more convenient to use the similarity criterion $D_N$, based on geocentric quantities, introduced by \citet{Valsecchi1999} for the identification of new meteor showers. The $D_N$ function is defined as follows:

\begin{equation}
{D_N}^2=[U_2-U_1]^2+[\cos{\theta_1}-\cos{\theta_2}]^2+\Delta\xi^2,
\label{eq:DN}
\end{equation}
where
\begin{equation}
\Delta\xi^2=\min(\Delta\phi_I^2+\Delta\lambda_I^2, \, \Delta\phi_{II}^2+\Delta\lambda_{II}^2)
\label{eq:xi}
\end{equation}

\begin{equation}
\Delta\phi_I=2\sin\left(\frac{\phi_1-\phi_2}{2}\right)
\label{eq:phi_Ii}
\end{equation}

\begin{equation}
\Delta\phi_{II}=2\sin\left(\frac{180^\circ+\phi_2-\phi_1}{2}\right)
\label{eq:phi_2}
\end{equation}

\begin{equation}
\Delta\lambda_I=2\sin\left(\frac{\lambda_1-\lambda_2}{2}\right)
\label{eq:lambda_I}
\end{equation}

\begin{equation}
\Delta\lambda_{II}=2\sin\left(\frac{180^\circ+\lambda_2-\lambda_1}{2}\right).
\label{eq:lambda_II}
\end{equation}

\noindent where $xyz$ is a Cartesian coordinate system centered on the Earth with the $y$ axis facing the direction of the Earth's motion, the $x$ axis opposite to the direction of the Sun and the $z$ axis orthogonal to the ecliptic plane, $D_N$ depends on 4 geocentric quantities directly related to the observations: $U$, the normalized geocentric speed of the body in units of the mean speed of the Earth (29.7 km s$^{-1}$); $\theta$, the angle between the direction of motion of the Earth along the $y$ axis and the $U$ vector; $\phi$, the angle between the $U$ vector and the $xz$ plane, and finally $\lambda$, the heliocentric longitude of the Earth at the encounter with the meteor shower. The angles $\theta$ and $\phi$ are directly related to the radiant, a quantity that for meteors and fireballs is directly observable and $U$ is a different way of considering the geocentric velocity of the meteoroid. Moreover, $D_N$ also has the advantage of being partially invariant under secular perturbations. The invariant quantities are $U$ and $\cos\theta$: the first because it depends only from the Tisserand invariant with respect to the Earth, the second because it depends from $U$ and $1/a$ which is a constant due to the conservation of orbital energy \citep{Valsecchi1999}. For the computation of $D_N$ it is assumed that the orbit of the Earth is circular with a radius equal to 1 au. In reality the orbit is slightly elliptical and this involves an error on $\theta$ and $\phi$ of the order of one degree, a negligible amount because the values of the cosine and sine functions change little. \\
For NEAs whose orbit intersects that of the Earth, the quantities $U$, $\theta$, $\phi$ and $\lambda$ that define the type of geocentric encounter can be computed once the heliocentric osculating orbital elements are known \citep{Valsecchi1999}. In the case of meteors or fireballs, the quantities $U$, $\theta$ and $\phi$ or the equivalent $\vec{U}=U_x\hat{\mathbf{x}}+U_y\hat{\mathbf{y}}+U_z\hat{\mathbf{z}}$ come from the geocentric position of the true radiant and the geocentric velocity according to the following formula \citep{Valsecchi1999} \footnote{There was a typo in \cite{Valsecchi1999}, the rotations are $-\lambda$ and $-\epsilon$.}:

\begin{equation}
\vec{U}= \hat{\mathbf{r}}(-\lambda) \hat{\mathbf{p}}(-\epsilon)\frac{V_G}{29.7}
\left(
\begin{array}{c}
-\cos\delta_G \cos\alpha_G \\
-\cos\delta_G \sin\alpha_G \\ 
-\sin\delta_G \\
\end{array}
\right).
\label{eq:U}
\end{equation}

\begin{figure*}
\centering
\includegraphics[width=\textwidth]{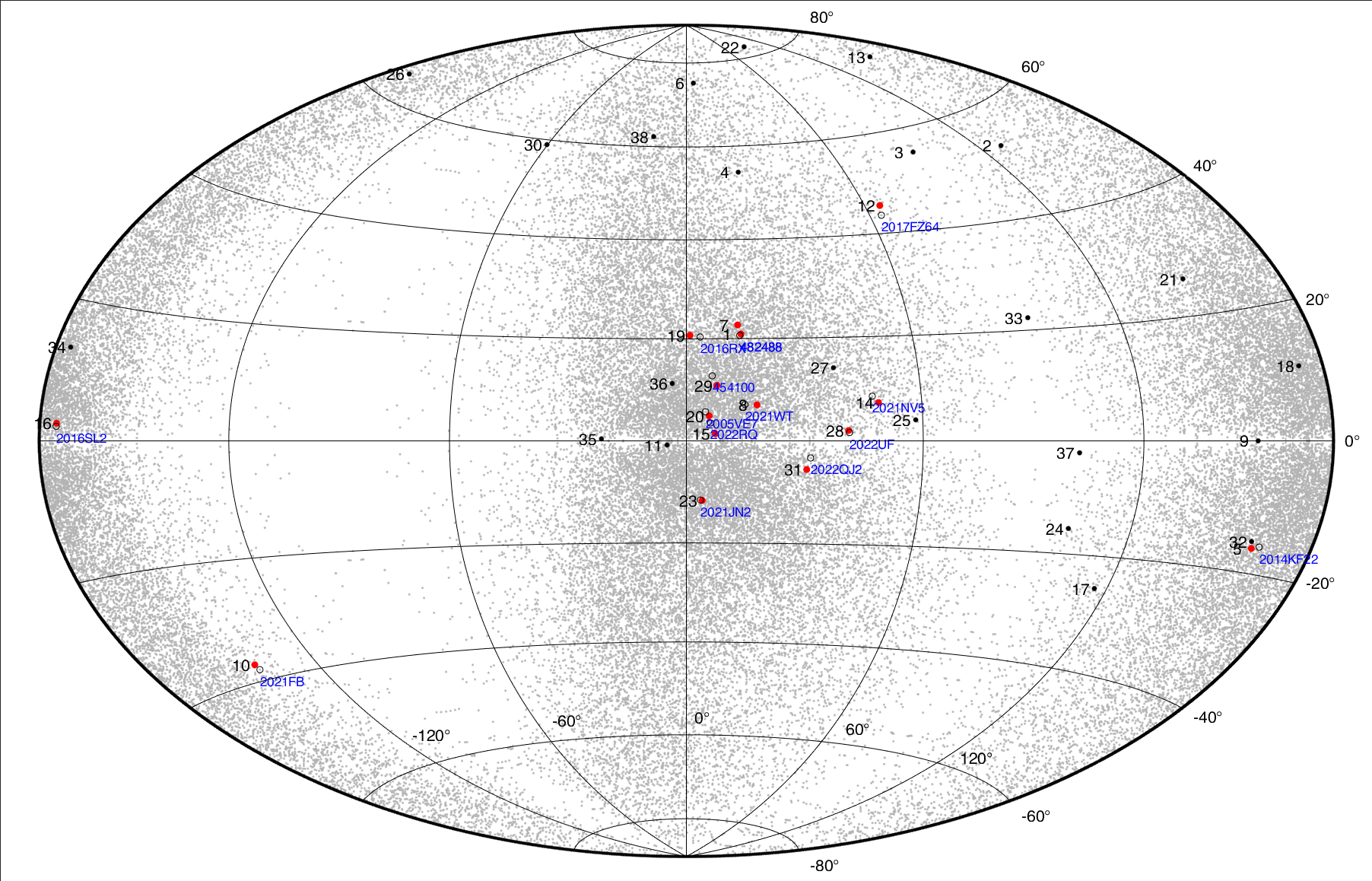}
\caption{The true geocentric all-sky radiants distribution, for each meteorite identified by the numbers listed in Table~\ref{tab:meteorites_encounters} in an equal area projection of the sky centered on the Sun opposition; the angular coordinates are ecliptic longitude minus the longitude of the Sun, and ecliptic latitude. The gray dots are the distribution of the NEAs radiants intersecting the Earth's orbit, the red dot are the meteorites with the candidate associated NEAs listed in Table~\ref{tab:ProgenitorCandidates}, the circles are the corresponding best candidate progenitor. The meteorites Innisfree (3), Sariçiçek (22), Murrili (24), Ejby (25) and Cavezzo (33), are found in areas where the density of NEAs radiants is low and the possible association with NEAs statistically simpler.}
\label{fig:meteorites_radiants}
\end{figure*}

In Eq. (\ref{eq:U}) $\hat{\mathbf{r}}(-\lambda)$ and $\hat{\mathbf{p}}(-\epsilon)$ are rotation matrices around $z$-axes and $x$-axes respectively, $\lambda$ is the ecliptic longitude of the Earth and $\epsilon$ is the inclination of the ecliptic above equatorial plane. The position of the true geocentric radiant $\alpha_G$ and $\delta_G$ is measured directly from the trajectory of the fireball, while to get the geocentric speed $V_G$ it is enough to correct the entry speed into the atmosphere for the gravitational attraction of the Earth: these are quantities directly correlated to the observables of the fireball. Unfortunately not all the papers relating to fireballs of which meteorites have been recovered and orbit computed have these data\footnote{Missing data for Almahata Sitta, Bouzzard Coulee, Mason Gully, Stubenberg, Tagish Lake and Viñales.}. We recommend that fireballs observers always include the true geocentric radiant and the geocentric velocity in their tables. Table~\ref{tab:meteorites_encounters}, computed from Eq. (\ref{eq:U}) with $\theta =\arccos{(U_y/U)}$ and $\phi=\arctan{(U_x/U_z)}$, shows $U$, $\theta$, $\phi$ and $\lambda$ values for the meteorites listed in Table~\ref{tab:meteorites}, excluding those for which information on true radian and geocentric velocity has not been found.\\
Figure~\ref{fig:meteorites_radiants} shows the distribution of the true geocentric radiants of the meteorites listed in Table~\ref{tab:meteorites_encounters} and those of NEAs that intersect the Earth's orbit. As we can see the meteorites Innisfree (3), Sariçiçek (22), Murrili (24), Ejby (25) and Cavezzo (33), are in areas where the density of NEAs is low and the association with a possible near-Earth is statistically simpler.

\subsection{The limit value for $D_N$}
\label{sec:DN} 
It is well known that the orbits of planet-crossing objects, including both NEAs and comets, are very chaotic so the orbital elements of NEAs evolve quickly due to the effect of planetary perturbations \citep{Gronchi2001}. In general, for the orbits of NEAs the most important gravitational perturbations can be secular, i.e. independent from the mean longitude on the orbit, resonant i.e. dependent by the semi-major axis of the orbit, or of short period if they cannot be classified in the first two types. Secular perturbations act on time scales of thousands of years and can be seen in the values of the argument of perihelion $\omega$ and of the ascending node $\Omega$ which can grow or decrease linearly over time through all the values between $0^\circ$ and $360^\circ$ and in those of the eccentricity and inclination which instead oscillates. On the other hand, the semi-major axis is invariant under secular perturbations. For a generic NEA that is not in resonance conditions, the most important perturbations to describe its long term orbital evolution are the secular ones. \\
For a typical NEA the secular $\omega$ cycle usually lasts from 5~000 to 10~000 years. For each $\omega$ cycle a NEA's orbit can intersect the Earth's orbit four times so they are called quadruple Earth crossers, but also octuple crossers are possible. Statistically, Earth-crossers are more probable in Apollo and Aten classes, defined on the basis of the current osculating elements, but Apollo and Aten are not equivalent to the class of Earth-crossing asteroids. Amor-type asteroids do not intersect the Earth's orbit, but a NEA can also switch from one orbital class to another with times of the order of thousands of years due to secular perturbations \citep{2001Icar..152...58G}. Considering that, for a given NEA, we do not know when the separation between the progenitor body and the meteoroid producing the meteorite took place, we must consider all the intersections that can occur with the Earth's orbit for one entire variation of $\omega$. In this way a connection can also be found between a meteorite and a NEA which are in a different phase of their $\omega$ cycle. 
For the secular evolution of NEAs we adopted the integration method by \citet{2001Icar..152...58G}, that allows to compute the secular evolution of the elements $e,i,\Omega$ and $\omega$ and to get all the crossing conditions.\\
The list of NEAs geocentric encounter conditions $U$, $\theta$, $\phi$ and $\lambda$ maintained by NEODyS-2 was used\footnote{\url{https://newton.spacedys.com/~neodys2/propneo/encounter.cond} This list were last consulted on Nov 18, 2022.}. This list was computed with the \texttt{OrbFit} software\footnote{\url{http://adams.dm.unipi.it/orbfit/}}
that gives the encounter conditions for 16~227 NEAs of the total 30~740 actual population which, during their $\omega$ cycle, intersect Earth's orbit. This is a crucial point and it should be emphasized: the computation of the encounters conditions for asteroids whose orbit is not crossing the Earth at the present time, but can cross at some time in the future and in the past, can be used in the search for parent bodies of meteor streams or fireballs \citep{2001Icar..152...58G}. A similar strategy can also be found in \citet{Babadzhanov2012}, who computes the geocentric radians of the orbits of NEAs crossing Earth's orbit on a $\omega$ cycle and compares them with those of observed meteor showers searching for dormant comets among the NEAs. As stated above a NEA can be a quadruple or octuple crossers, depending on the number of possible intersections with the Earth's orbit during the $\omega$ cycle. The intersection can take place at the ascending or descending node and - fixed the node - can be pre or post perihelion.\\
Once you have a criterion definition, it is also necessary to establish the maximum $D_N$ value ($D_{N_{\max}}$), beyond which the association between meteorites and NEAs begins to lose meaning due to the occurrence of random associations. For this purpose, two populations of NEAs were considered: observed and randomized, both evolved in $\omega$. As the value of $D_N$ increases starting from low number, it is expected that the number of randomized NEAs associated with the observed meteorites will increase slower than the population of observed NEAs, because with the randomization we have erased the underlying evolution, if any. As $D_N$ increases further, there will be a tendency to have the same number of associations between the real and random populations because the high value of the distance will make the spurious associations ever more numerous. In this way it is possible to estimate the maximum value of $D_N$ above which the physical association NEA$-$meteorite begins to lose its meaning due to the large number of random associations \citep{Rozek2011}.\\
The population of randomized NEAs evolved into $\omega$ was constructed starting from the observed one and considering $U$, $\theta$ and $\phi$ as independent variables and randomly adding the probability density of the analogous quantities present in the NEAs observed population intersecting the Earth in their $\omega$ cycle. The longitude of the Earth $\lambda$ instead has been replaced by random values taken between 0 and 360 degrees, which is equivalent to randomizing the date of the encounter with the Earth.\\
Using the meteorites encounter conditions listed in Table~\ref{tab:meteorites_encounters} and computing $D_N$ with both observed and random NEAs using Eq.~\eqref{eq:DN}, we see that the number of possible meteorite$-$NEA pairs for small $D_N$ values rises much faster for the real population than for the virtual one, while for high values of $D_N$ the number of pairs tends to be similar, see Table~\ref{tab:pairs-real-virtual}. By generating different virtual populations, there may be random fluctuations in the number of meteorite$-$NEAs pairs for small values of $D_N$, an expected behavior because the population consists of tens of thousands of virtual asteroids which may be closer to a meteorite's radiant than the real population. The value beyond which the number of meteorite$-$NEA pairs for the real population suddenly rises and stable spurious associations start to appear is 0.05, a value that we can take as a limit. However, we took a slightly larger $D_N$ value of 0.06, to be sure not to miss any potential meteorite-NEA pairs. In our case it will be the subsequent dynamic analysis of the orbits to establish the possible physical connection, so $D_{N_{\max}} = 0.06$. The values reported in Table~\ref{tab:pairs-real-virtual} refer to a single virtual population, and we can see that they are affected by small number statistics: trying to extrapolate an upper $D_N$ limit by fixing an exact small probability for a spurious association may not give an appropriate result. The choice of $D_{N_{\max}} = 0.06$ is a safer and conservative choice, and it is also consistent with the values found by \citet{Jopek1999} for the similarity of pairs and triplets about 865 meteor radiants with $D_{N_{\max}}\approx 0.02-0.08$.

\begin{table}
	\centering
	\caption{Number of meteorite$-$NEA pairs for observed (Obs) and randomized (Rand) NEAs populations evolved in $\omega$. In the last column the probability of having a true meteorite-NEA pair is estimated. As we see, spurious pairs start to appear for $D_N \geq 0.05$ but we took a slightly larger $D_{N_{\max}} = 0.06$ value to be sure not to miss any potential meteorite-NEA pairs.}
	\label{tab:pairs-real-virtual}
	\begin{tabular}{lccc} 
		\hline
		$D_{N_{\max}}$ & $\text{Obs}$ & $\text{Rand}$ & $(\text{Obs}-\text{Rand})/\text{Obs}$\\
		\hline
0.010       &    0   &   0 & 1.00\\
0.020       &    0   &   0 & 1.00\\
0.030       &    1   &   0 & 1.00\\
0.040       &    1   &   0 & 1.00\\
0.050       &    12  &   2 & 0.83\\
0.055       &    16  &   2 & 0.87 \\
0.056       &    18  &   4 & 0.78 \\
0.057       &    19  &   4 & 0.79\\
0.058       &    21  &   4 & 0.81\\
0.059       &    22  &   4 & 0.82 \\
0.060       &    22  &   4 & 0.82\\
0.061       &    22  &   6 & 0.73\\
0.062       &    25  &   6 & 0.76\\
0.063       &    25  &   6 & 0.76\\
0.064       &    29  &   6 & 0.79\\
0.065       &    30  &   6 & 0.80\\
0.070       &    41  &  10 & 0.76\\
0.080       &    66  &  16 & 0.76\\
0.090       &    98  &  35 & 0.64\\
0.100       &   141  &  51 & 0.64\\
\hline
	\end{tabular}
\end{table}

\section{Association between NEAs and meteorites}
\label{sec:association}
Now that we have established a maximum value of $D_N$ beyond which it is not advisable to go, from Eq.~\eqref{eq:DN} we can compute $D_N \leq 0.060$ from the meteorites of Table~\ref{tab:meteorites_encounters} with the NEAs population $\omega$-evolved and we found the possible meteorite-NEAs associations given in Table~\ref{tab:ProgenitorCandidates}: in total there are 20 possible meteorite$-$NEA pairs. \\
These candidate NEAs are mostly small, with diameters of a few tens of meters although there are also objects up to 1 km in diameter, and none of them have been physically characterized. 
The asteroids that have been observed from the photometric point of view are 482488, 2005 VE7 and 454100. The first two appear in the catalog of Solar System Objects extracted from the Sloan Digital Sky Survey \citep{Sergeyev2021}. This catalog contains data on 1~032~357 Solar System objects (asteroids, Kuiper Belt Objects, and comets) that were classified in a scheme consistent with Bus-DeMeo taxonomy.\\ 
Asteroid 482488 is classified as U (unknown type), while for 2005 VE7 there are two distinct observations. In the first one, it has a $67\%$ probability of being an S-type, in the second one it is classified as U. Asteroid (454100) 2013 BO73 is listed in the Asteroid Lightcurve Database \citep{Warner2009} with a rotation period of about 1.1 h. Unfortunately, the photometric observations are incomplete and it is possible that the period is completely wrong. If confirmed, it would be a value well below the cohesionless spin-barrier value valid for rocky asteroids  - about 2.2 h, see \cite{Carbognani2017} - and 454100 would be a natural candidate for rotational disruption. So of all the asteroids in Table~\ref{tab:ProgenitorCandidates} only 2005 VE7 has a probability of being an S-type. For the others asteroids there are not useful information.\\ 
According to the Tisserand parameter with respect to Jupiter, all the progenitor candidates have asteroidal-type orbits except 2016 RX, which is just under three. In general, the orbits of meteorites in Table~\ref{tab:ProgenitorCandidates} tend to have a low inclination on the ecliptic except Pribram, Neuschwanstein, Bunburra, Jesenice and Annama which have inclinations higher than about $9^{\circ}$. At first glance, the most interesting result is that Pribram and Neuschwanstein have the same NEA as a possible progenitor. This is not surprising considering that these two meteorites have a very similar orbit, a fact that was immediately noticed after the Neuschwanstein event \citep{Spurny2003}. Note that there are three meteorites, Hamburg, Košice and Križevci which are potentially associated with two NEAs at the same time, while Motopi Pan is associated with three NEAs. The best progenitor candidate of the meteoroid that produced Motopi Pan, the asteroid 454100, was also found by \cite{deLaFuente2019} using D$-$criteria.\\
The asteroids in Table~\ref{tab:ProgenitorCandidates} have low escape velocities, in the range 0.01$-$1 m s$^{-1}$ if we assume the typical average density for S-type asteroids of $2.72\pm 0.54~\textrm{g }\textrm{cm}^{-3}$ \citep{Carry2012}. Therefore any physical mechanism invoked to extract a meteoroid from their surface must be able to impart at least a velocity higher than this range. In the case of an impact, statistically the larger ejecta have a slower velocity than the smaller ones which tend to travel further away from the impact site. The velocity of the ejecta falls approximately in the range $V\sim 10^{-4}-10^{-1} ~U_i$, where $U_i$ is the impact speed \citep{Ferrari2022}. Considering that the typical collision velocity between NEAs is of the order of 10$-$15 km s$^{-1}$ \citep{Bottke1993}, we can expect ejecta velocities in the range 1.0 m s$^{-1}$ $-$ 1.5 km s$^{-1}$ or at most of few km per second. About the rubble-pile rotational instability, by definition the expulsion occurs when the meteoroid on the NEA's surface exceeds the escape velocity \citep{Jewitt2012}, while for tidal destruction by the Earth the relative velocity between the fragments can be estimated in the order of 1 m s$^{-1}$ \citep{Schunova2014}. Finally, thermal fracturing of the surface rocks can lead to ejection speeds of the order of 20 m s$^{-1}$ for a $\Delta T$ of about 1000 K \citep{Jewitt2012}.
All these speed values are equal or higher than the escape speeds of the NEAs in Table~\ref{tab:ProgenitorCandidates}, therefore all these mechanisms are potentially able to cause a spatial separation between a NEA and its components. The ejection velocities are much smaller than NEAs orbital speed, so the orbits of the ejected meteoroid will be similar to that of the parent body. Only in the case of collisions, one can expect an ejection velocity much higher than the escape velocities, so in this case it is reasonable to expect a faster divergence between the orbits than in the other - less energetic - cases.\\
To verify if the candidate pairs in Table~\ref{tab:ProgenitorCandidates} have a real physical meaning, we checked the past orbital evolution of both meteorite and NEA. In fact, even if we chose $D_{N_{\max}}$ in such a way to avoid random associations as much as possible, the large number of NEAs and meteorites with low orbital inclination could make the association random anyway.

\begin{table}
    \centering
    \caption{List of meteorites with an associated NEA with $D_{N}\leq 0.06$. The column with Diam. is the estimated diameter, $T_J$ is the Tisserand invariant with respect to Jupiter ($T_J < 3$ indicate a comet-like orbit), while the column with $C$ is a value from the NEOCC database between 0 and 9 indicating how well an object's orbit is known on a logarithmic scale, where 0 indicates a well-determined orbit.}
    \label{tab:ProgenitorCandidates}
    \begin{tabular}{lccccc} 
    \hline
    Meteorite Name & NEA & Diam. (m) & $T_J$ & $C$ & $D_N$  \\
    \hline
    Pribram			      &	482488	 & 300$-$600	& 3.109	&	0.3 &   0.047 \\
    Peekskill		      &	2014 KF22 & 15$-$30	& 4.444	&	5.9 &   0.041 \\
    Neuschwanstein	      &	482488	 & 300$-$600	& 3.109	&	0.3 &   0.056 \\
    Park Forest			  & 2021 WT   & 30$-$60    & 3.178 &   7.6 &   0.053 \\
    Bunburra Rockhole     & 2021 FB	 & 20$-$40    & 6.942 &   6.4 &   0.042 \\
    Jesenice			  & 2017 FZ64 & 40$-$100   & 3.736 &   7.8 &   0.055 \\
    Ko\v sice			      &	2021 NV5	 & 10$-$20	& 3.234	&	6.8 &   0.049 \\
                          & 2019 ST2  & 50$-$120   & 3.288 &   9.0 &   0.059 \\
    Kri\v zevci		      &	2022 RQ	 & 20$-$50	& 4.472	&	7.9 &   0.044 \\
                          & 2013 BR15 & 30$-$60    & 4.280 &   9.0 &   0.056\\
    Sutter's Mill		  & 2016 SL2  & 30$-$60    & 3.327 &   8.7 &   0.050 \\
    Annama				  & 2016 RX	 & 20$-$40    & 2.943 &   7.1 &   0.048 \\
    \v Zd'ár nad Sázavou     &	2005 VE7	 & 500$-$1200	& 3.156	&	0.5 &   0.048 \\
    Creston				  & 2021 JN2	 & 40$-$100	& 5.328 &   7.0 &   0.055 \\
    Hamburg 			  &	2022 UF	 & 10$-$20	& 3.110	&	6.8 &   0.057 \\
                          & 2021 PZ1  & 20$-$40  & 3.279 &   7.6 &   0.043\\
    Motopi Pan            & 454100   & 500$-$600 & 4.824 &  0   & 0.044  \\
                          & 2017 MC3  & 40$-$100 & 4.333 &  3.5   & 0.046  \\
                          & 2009 FZ4  & 20$-$50  & 5.535 &  2.0   & 0.055  \\
    Arpu Kuilpu           & 2022 QJ2  & 20$-$50  & 3.067 &  8.2   & 0.057  \\
    \hline
    \end{tabular}
\end{table}

\section{Orbits evolution}

\label{sec:integration}
Numerical integrations were performed with the \texttt{mercury}\footnote{\url{https://github.com/Fenu24/mercury}} $N$-body code \citep{Chambers1997}, by using the Bulirsch-Stoer integration method \citep{Stoer2002}. The dynamical model used for the propagation of the orbits was purely gravitational, and include the attraction of the Sun and the perturbations of the eight planets from Mercury to Neptune. Initial conditions of the planets at the epoch of a meteorite fall were computed with the \texttt{OrbFit} software, that includes a planetary ephemerides generator based on the JPL DE431 ephemerides system \citep{Folkner2014}.

Uncertainties of meteorite orbits are generally larger than those of NEAs orbits, because meteorites are observed only in a really short interval of time during the fall. To take into account such uncertainties and provide a deeper analysis of the correlation between a meteorite and a potential progenitor NEA, we produced orbital clones of the meteorite. To this purpose, we assumed that each orbital element is Gaussian distributed, with a mean value corresponding to the nominal value and a standard deviation equal to the $1\sigma$ uncertainty (see Table~\ref{tab:meteorites}). \\
The mean anomaly $M$ was computed by using the Kepler equation with the true anomaly $f$ at the epoch of the fall, that is determined only by the geometry of the orbit and the position along it as follows:

\begin{equation}
    f = 
    \begin{cases}
        -\omega            & \text{asc. node fall, $\omega < 180^\circ$},\\
        180^\circ - \omega & \text{des. node fall, $\omega < 180^\circ$},\\
        360^\circ - \omega & \text{asc. node fall, $\omega > 180^\circ$},\\
        180^\circ - \omega & \text{des. node fall, $\omega > 180^\circ$}.\\
    \end{cases}
    \label{eq:TrueAnomaly}
\end{equation}

Equation \eqref{eq:TrueAnomaly} is a geometric relation between the true anomaly $f$ at the time of passage through the ascending or descending node, and the argument of perihelion $\omega$ that does not involve the semi-major axis of the orbit. Note that the fall necessarily happens at the passage through the node. As a test, we verified that the value of the mean anomaly $M$ for Almahata Sitta (due to the fall of asteroid 2008 TC3) and Motopi Pan (due to the fall of 2018 LA), see Table~\ref{tab:meteorites}, was in good agreement with the value given by JPL Small- Body Database\footnote{\url{https://ssd.jpl.nasa.gov/}}.
The $1\sigma$ uncertainty in the mean anomaly $M$ was assumed to be the same as the $1\sigma$ uncertainty in $\omega$. 

For the meteorites of Table~\ref{tab:ProgenitorCandidates} with an identified candidate progenitor, we simulated the backward evolution of 5000 orbital clones. The orbits were propagated from the epoch of the fall to 100 kyr prior the fall, by using a maximum timestep of 0.5 days. Output orbital elements were recorded every 50 yr. The nominal orbit of the NEA progenitor was also integrated backwards for the same timespan and with the same dynamical model. Keplerian initial conditions of NEAs of Table~\ref{tab:ProgenitorCandidates} at epoch 2460000.5 JD were taken for the NEOCC portal\footnote{\url{https://neo.ssa.esa.int/}}, and are also reported in Table~\ref{tab:neo_orbits}.

\subsection{Pribram$-$482488 and Neuschwanstein$-$482488}
\label{ss:482488}
\begin{figure*}
    \centering
    \includegraphics[width=\textwidth]{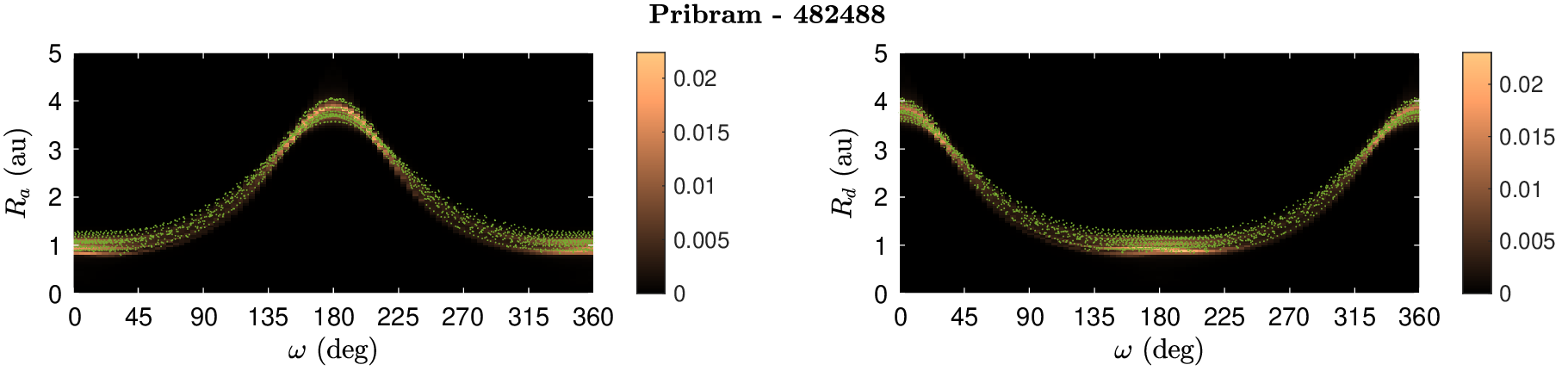}
    \includegraphics[width=\textwidth]{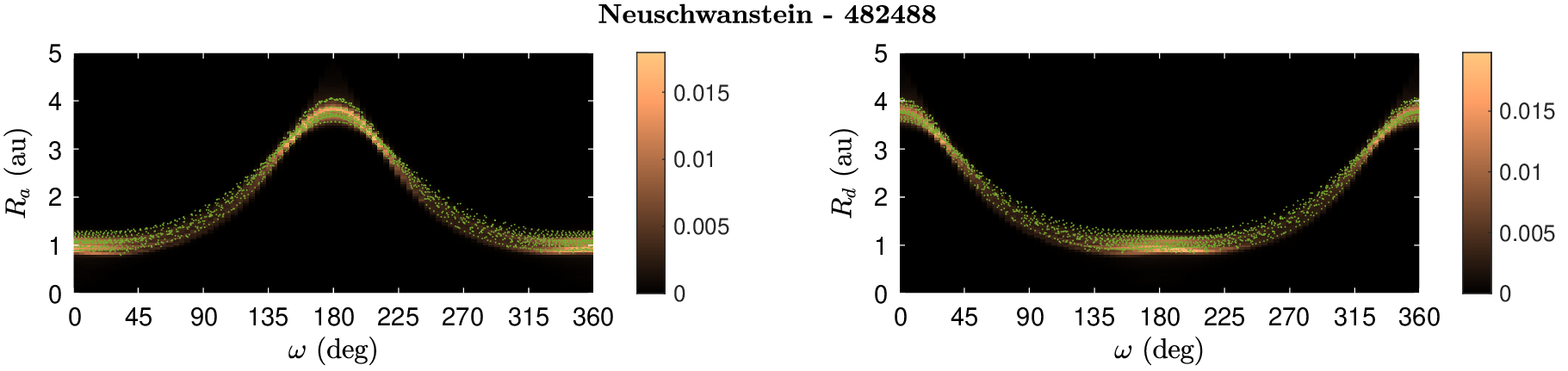}
    \caption{Evolution in the planes $(\omega, R_a)$ (left column) and $(\omega, R_d)$ (right column), for the pairs Pribram$-$482488 (first row) and Neuschwanstein$-$482488 (second row). The green dots denote the evolution of the NEA parent body 482488. The evolution of the meteorite orbital clones is shown as a density plot, computed over the whole 5000 orbits. }
    \label{fig:Pribram_Neusch_482488}
\end{figure*}

We first analyze the possible correlation between the Pribram and Neuschwanstein meteorites and the NEA 482488. %
\citet{Babadzhanov2012} identified the meteor showers associated with 2004CK39 by taking as an indicator the time evolution in the planes $(\omega, R_a)$ and $(\omega, R_d)$, where 
\begin{equation}
    R_a = \frac{a(1-e^2)}{1+e\cos\omega}, \\ 
    R_d = \frac{a(1-e^2)}{1-e\cos\omega},
    \label{eq:NodalDistances}
\end{equation}
are the ascending and descending node distances. This is done because the time evolution of $R_a, R_d$ could be very different between the meteorite and the candidate NEA progenitor. In \citet{Babadzhanov2012}, the authors found similar evolutionary paths for the NEA and the meteor showers, and they therefore suggested a possible correlation. We followed a similar approach to provide an evidence of the common origin of the Pribram and the Neuschwanstein meteorites through the NEA 482488. Because the backward evolution of a meteorite orbit is propagated using orbital clones, we took into account the density distribution in the planes $(\omega, R_a)$ and $(\omega, R_d)$, computed over the whole 5000 orbits. Figure~\ref{fig:Pribram_Neusch_482488} shows the density plots for the two pairs Pribram$-$482488 and Neuschwanstein$-$482488. The path of the nominal orbit of the NEA, depicted with green dots, is superimposed on the density plots. In both cases, the evolution of 482488 follows almost exactly the areas with the largest density, denoted by the orange color. This already provides a first indication of the common origin of these objects. 
\begin{figure*}
    \centering
    \includegraphics[width=0.48\textwidth]{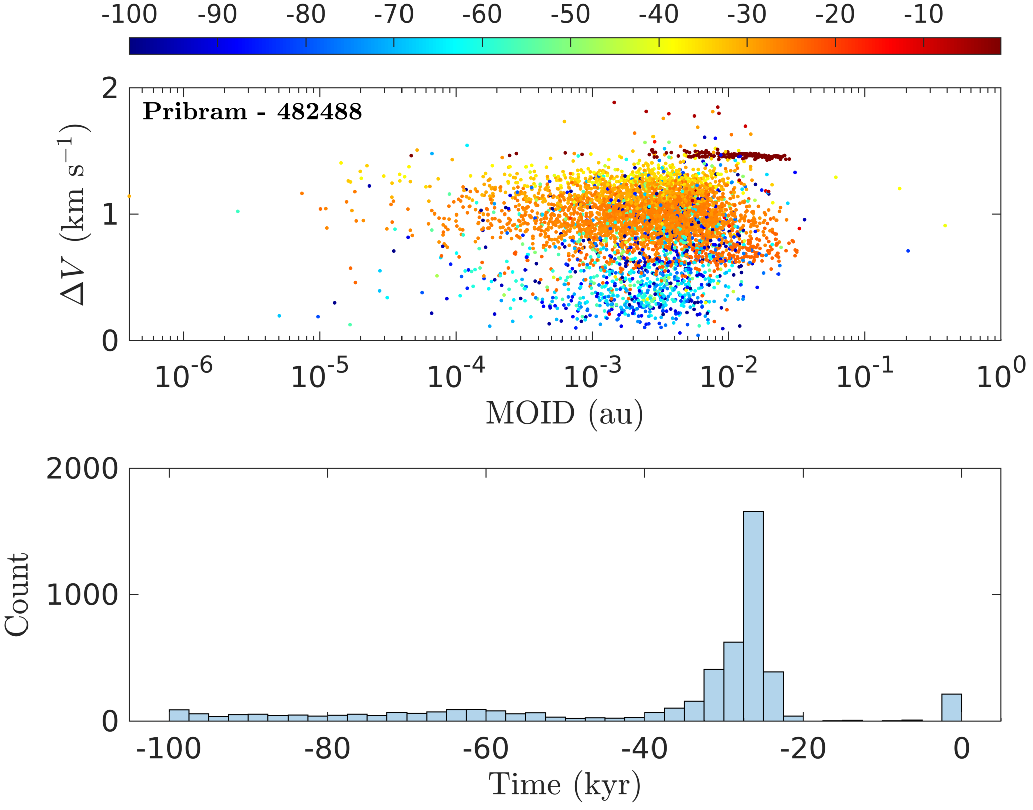}
    \includegraphics[width=0.48\textwidth]{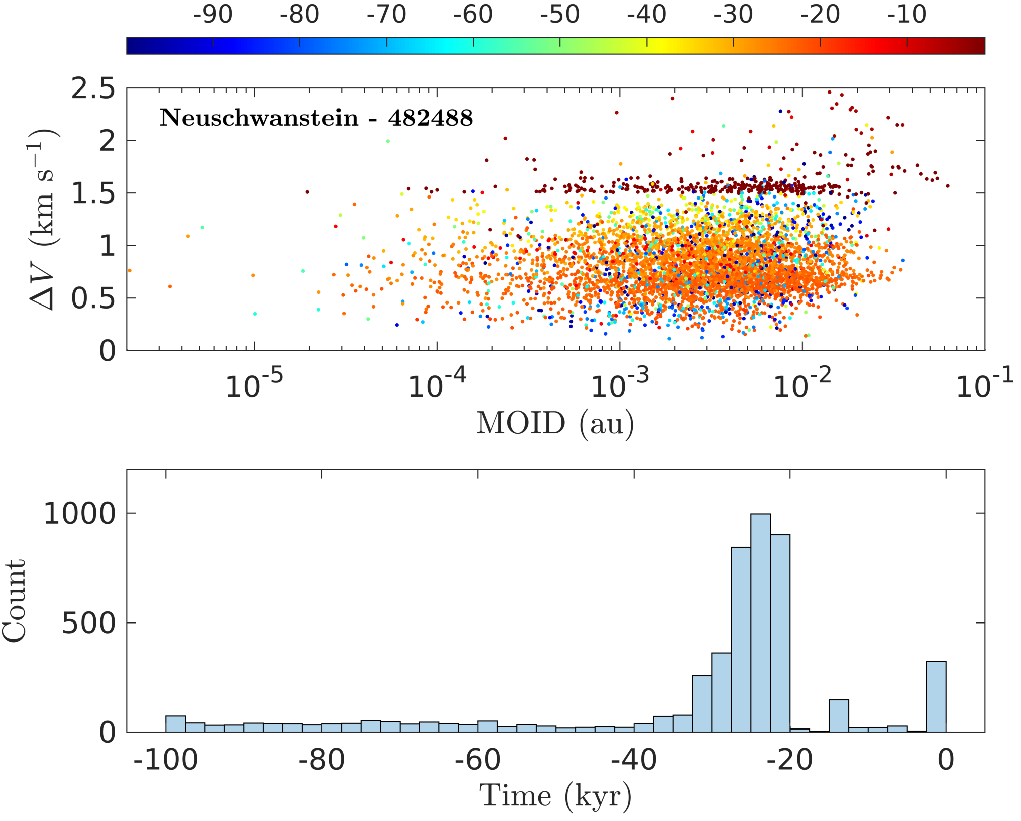}
    \caption{Distributions of the relative velocity vs. the MOID at the time $t_{\min}$ (first row), and histogram of $t_{\min}$ (second row).} Plots in the left column refer to the pair Pribram$-$482488, while those of the right column to Neuschwanstein$-$482488. The unit for the colorbar in the first row is kyr. The bins in the histograms of the second row have a constant width of 2.5 kyr.
    \label{fig:Pribram_Neusch_mindv_vs_moid}
\end{figure*}

To understand whether it is possible to determine a candidate age for the separation between the meteorite and the proposed parent body, we searched for convergence events involving a small relative velocity and a small distance between the two objects. To this purpose, we adopted an approach similar to that used by \citet{Moskovitz2019} to study the two NEAs asteroid pairs 2015 EE7$-$2015 FP124 and 2017 SN16$-$2018 RY7. The distance was evaluated in terms of the Minimum Orbit Intersection Distance (MOID) between the orbit of the meteorites and that of 482488, while for the velocity we considered the norm of the difference between the heliocentric velocities. Since the errors in the initial orbital elements are typically accumulated in the mean anomaly in long-term numerical propagation, the instantaneous relative velocity is not very informative for the purpose of finding a possible separation age. Instead, we considered the relative velocity $\Delta V$ at the mean anomaly that realizes the MOID. At the separation event, we expect both the MOID and the relative velocity to be small. To find such events, we computed the dimensionless distance
\begin{equation}
    d = \sqrt{
 \bigg(   \frac{\text{MOID}}{\mu(\text{MOID})} \bigg)^2+ 
 \bigg(   \frac{\Delta V}{\mu(\Delta V)} \bigg)^2,
 \label{eq:distAlbino}
} 
\end{equation}
where $\mu(\text{MOID})$ and $\mu(\Delta V)$ are the median values of the MOID and the relative velocity at the MOID, respectively, computed over the whole integration timespan. For each clone, we recorded the minimum $d_{\min}$ of the distance $d$ and the corresponding time $t_{\min}$, together with the values of the MOID and $\Delta V$ attained at $t_{\min}$. 

Figure~\ref{fig:Pribram_Neusch_mindv_vs_moid} shows the distribution of the MOID and $\Delta V$ (first row) and that of the time $t_{\min}$ (second row). Plots on the left column refer to the pair Pribram$-$482488, while those on the right column to Neuschwanstein$-$482488. In both cases, we can note accumulations of the MOID at values smaller than 0.1 au during the first $5$ kyr of backwards evolution, happening at a relative velocity of $\sim$1.5 km s$^{-1}$. In addition, a large concentration of minima are found for MOID between 10$^{-4}$ au and 10$^{-5}$ au at relative velocities of $\sim$1 km s$^{-1}$ for the pair Pribram$-$482488, and smaller than $\sim$1 km s$^{-1}$ for Neuschwanstein$-$482488. Some isolated points of the distribution extend even towards MOID $\sim 10^{-5}$ au and $\Delta V \sim 0.5$ km s$^{-1}$. These large concentrations happen mostly between $-25$ and $-30$ kyr for Pribram$-$482488 and between $-20$ and $-25$ kyr for Neuschwanstein$-$482488. Beyond $-40$ kyr of evolution we cannot distinguish anymore any clear concentrations of small MOID and small $\Delta V$, and we believe this to be caused by both the chaotic nature and the uncertainties of the orbits of the meteorites, and an identification of a separation event dating further back is not possible in these cases.
Even if the relative velocities did not reach very low values of the order of few tens of m s$^{-1}$ as found in the NEAs pairs by \cite{Moskovitz2019}, they still may suggest that the Pribram and Neuschwanstein both originated from 482488, and the peaks in the distributions of $t_{\min}$ would suggest a separation age between $-20$ and $-30$ kyr. \\
These are the results that can be obtained from a dynamical point of view. If we consider the cosmic-ray exposure age of the two meteorites we see that they differ: for Pribram it is estimated in 12 Myr, while for Neuschwanstein it is around 48 Myr \citep{Spurny2003}. However, the cosmic-ray exposure time of a meteorite does not necessarily coincide with the time spent in space by the original meteoroid, rather it is the time period in which the meteorite was bombarded by cosmic rays, which can also occur up to about one meter below the surface of the progenitor body \citep{Eugster2003}. This condition can occur if the progenitor NEA is, for example, a rubble pile, as can be 482488. In this case it is reasonable to expect that the asteroid is composed of fragments with a not coincident cosmic-ray exposure history.

\subsubsection{Influence of the Yarkovsky effect}
For these two pairs of meteorite$-$NEA we performed additional simulations that included the Yarkovsky effect in the orbital evolution. The pre-atmospheric mass of Neuschwanstein was estimated to be about $300 \pm 100$ kg \citep{Spurny2003}, corresponding to a diameter of about 0.55 meters assuming a measured mean density $\rho \approx 3.5$ g cm$^{-3}$ \citep{Flynn2018}. It is reasonable to assume that Pribram was also similar in size and mass. To estimate the semi-major axis drift caused by the Yarkovsky effect we used the calibration based on Bennu \citep{Spoto2015} given by
\begin{equation}
    \bigg( \frac{\text{d}a}{\text{d}t} \bigg) = \bigg( \frac{\text{d}a}{\text{d}t} \bigg)_{\text{B}} \times \frac{\sqrt{a_{\text{B}}}(1-e_{\text{B}}^2)}{\sqrt{a}(1-e^2)} \frac{D_{\text{B}}}{D} \frac{\rho_{\text{B}}}{\rho} \frac{\cos \gamma}{\cos\gamma_{\text{B}}} \frac{1-A}{1-A_{\text{B}}}, 
    \label{eq:yarkovskyCalibration}
\end{equation}
where $\rho$ is the density, $D$ is the diameter, $\gamma$ is the obliquity, and $A$ is the Bond albedo. The subscript B denotes the values corresponding to asteroid Bennu, and a summary of the numerical values can be found for instance in \citet{DelVigna2018}. For the meteorite, we used $\rho = 3.5$ g cm$^{-3}$, $D = 0.55$ m and $\gamma = 180^\circ$, obtaining a semi-major axis drift of $\text{d}a/\text{d}t = -0.64$ au My$^{-1}$, that we maintained constant through the whole integration time-span. Numerical integrations were performed with a modified version of the \texttt{mercury} integrator \citep{Fenucci2022b}, by using the same orbital clones used for the previous simulations. The evolution in the planes $(\omega, R_a)$ and $(\omega, R_d)$ was not affected by the introduction of the Yarkovsky effect in the model, and therefore it is not reported.
\begin{figure*}
    \centering
    \includegraphics[width=0.48\textwidth]{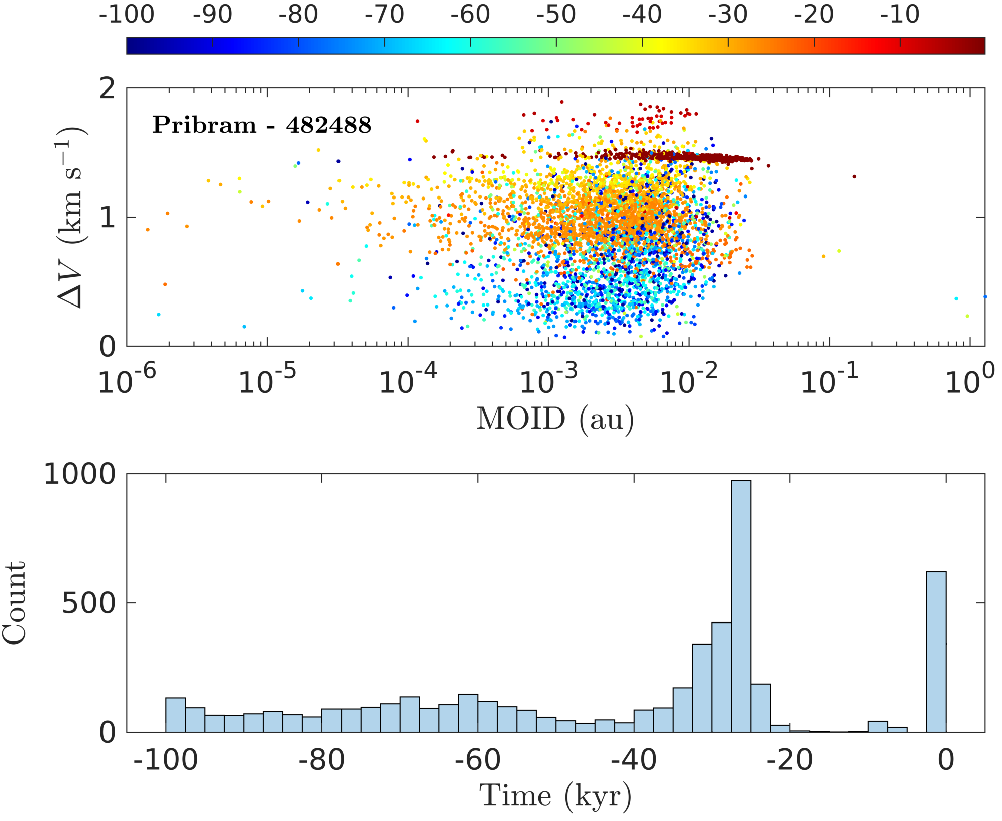}
    \includegraphics[width=0.48\textwidth]{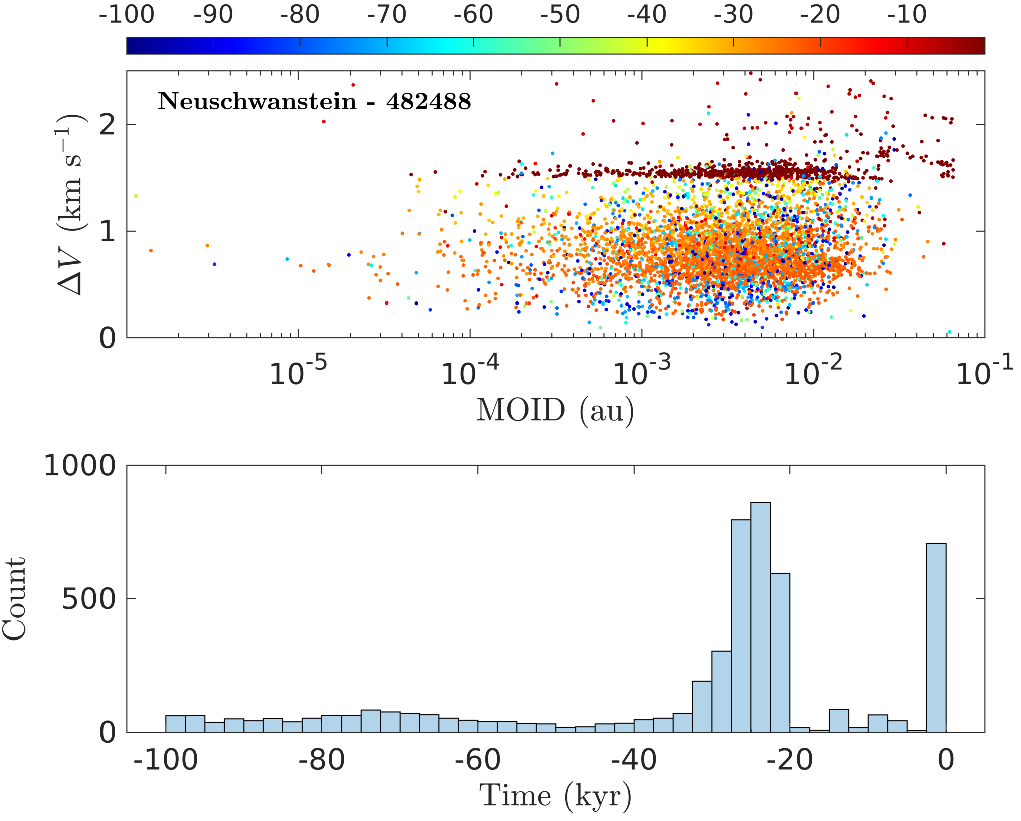}
    \caption{Same as Fig.~\ref{fig:Pribram_Neusch_482488} for the simulations performed by including the Yarkovsky effect in the model.}
    \label{fig:Pribram_Neusch_mindv_vs_moid_yarkovsky}
\end{figure*}

On the other hand, both the distributions of the values of the MOID and the relative velocities $\Delta V$ at the minima of $d$, and that of $t_{\min}$, show differences with respect to those obtained in a purely gravitational model (see Fig.~\ref{fig:Pribram_Neusch_mindv_vs_moid_yarkovsky}). A larger concentration of points at velocities $\Delta V \sim 1.5$ km s$^{-1}$ can be seen in the simulations that included the Yarkovsky effect in the model, happening during the first 2.5 kyr of backward evolution, thus producing a new more prominent peak in the distribution of $t_{\min}$. On the other hand, the peaks between $-20$ and $-30$ kyr are still present, suggesting that the proximity result between the orbits of the meteorites and that of 482488 is robust. It is important to note that the semi-major axis drift estimate obtained by Eq.~\eqref{eq:yarkovskyCalibration} is the maximum drift that can be obtained, because we fixed the obliquity to $180^\circ$. However, the Yarkovsky effect may result to be much smaller for different reasons. A first reason is that the diurnal Yarkovsky effect is proportional to $\cos\gamma$, hence this component is not effective at obliquity equal to $90^\circ$. A second reason is that the rotation period of small asteroids can reach very small values of the order of few seconds \citep{Pravec2000}, causing the thermal re-emission to be isotropic and hence slowing down the net Yarkovsky effect. Finally, the Yarkovsky–O’Keefe–Radzievskii–
Paddack (YORP) effect \citep{Rubincam2000, Bottke2006} is another thermal effect strictly related to the Yarkovsky effect, that causes a time evolution of both the obliquity and the rotation period. The magnitude of this effect scales as $1/D^2$, and the YORP cycle (i.e. the time needed to reach one of the asymptotic obliquity values of $0^\circ, 90^\circ$ or $180^\circ$) can be very short for meter-sized objects. This causes a random walk in the semi-major axis drift $\text{d}a/\text{d}t$ \citep{Bottke2015}, thus decreasing the total average drift. The simulation presented here represents the most extreme case, in which the Yarkovsky effect is maximum and it is kept at the same value for the whole duration of the integration timespan, hence it is reasonable to assume that the results we obtained in the purely gravitational case are stable also in the case of smaller semi-major axis drifts.

In the following sections we use similar methods to those used here to discuss the possible confirmation of the meteorites$-$NEAs pairs of Table~\ref{tab:ProgenitorCandidates}, except for the inclusion of the Yarkovsky effect in the model.

\subsection{Annama$-$2016 RX}
\label{ss:annama}
\begin{figure*}
    \centering
    \includegraphics[width=\textwidth]{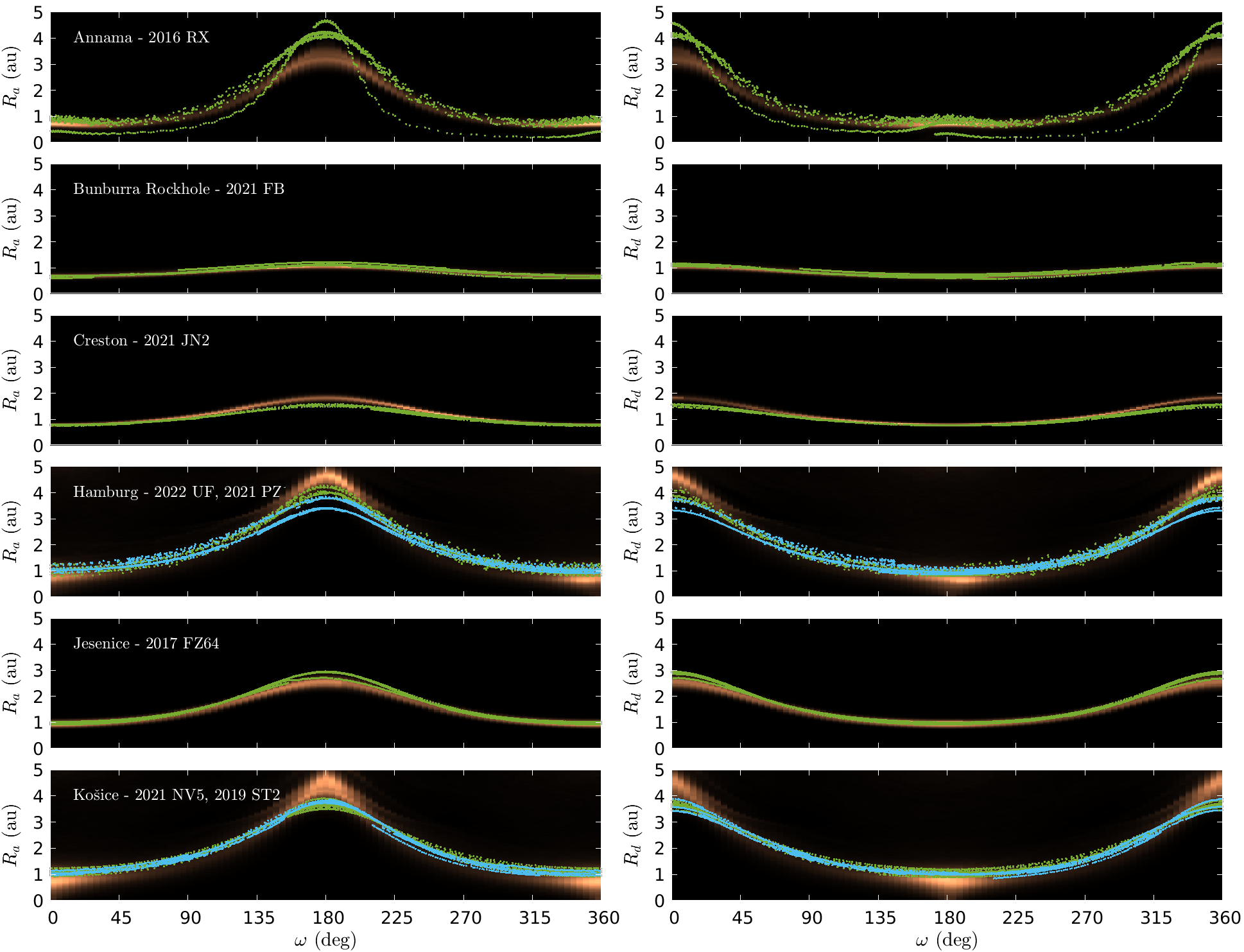}
    \caption{Same as Fig.~\ref{fig:Pribram_Neusch_482488} for the pairs Annama$-$2016 RX (first row), Bunburra Rockhole$-$2021 FB (second row), Creston$-$2021 JN2 (third row), Hamburg$-$2022 UF, 2021 PZ1 (fourth row), Jesenice$-$2017 FZ64 (fifth row), and Košice$-$2021 NV5, 2019 ST2 (sixth row). The green color represents the evolution of the primary associated NEA, and the cyan color the evolution of the secondary associated NEA, when present.}
    \label{fig:5_met}
\end{figure*}

\begin{figure}
    \centering
    \includegraphics[width=0.48\textwidth]{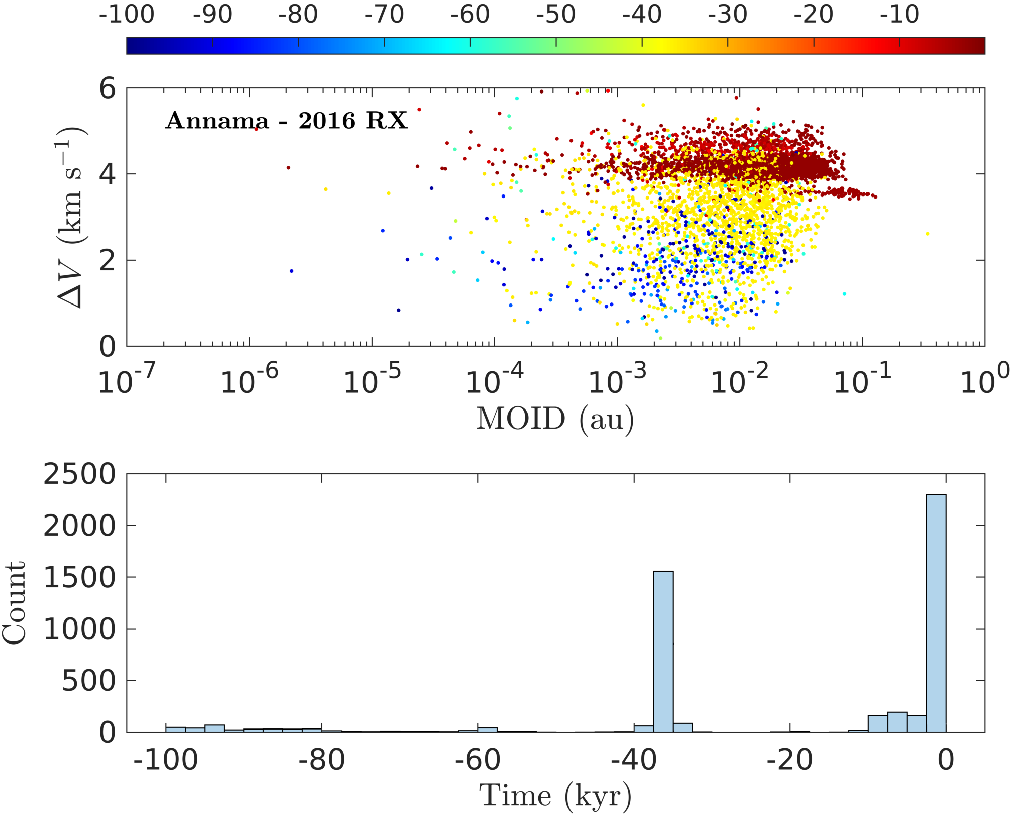}
    \vskip 10pt
    \includegraphics[width=0.48\textwidth]{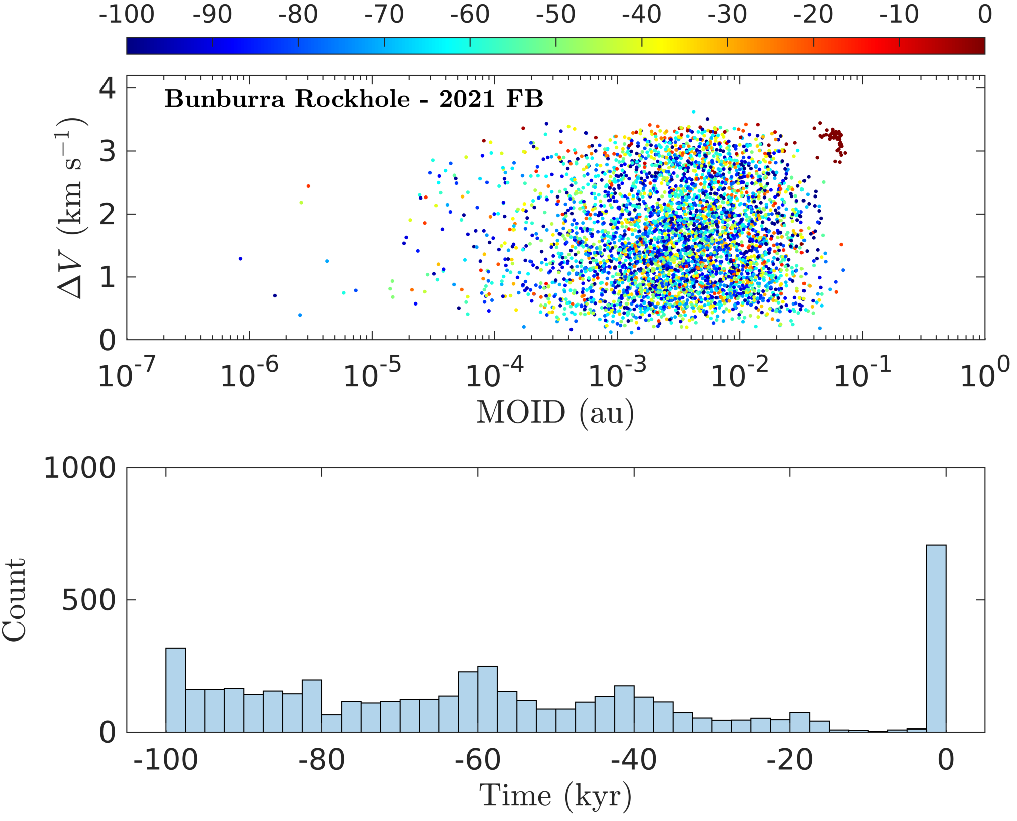}
    \vskip 10pt
    \includegraphics[width=0.48\textwidth]{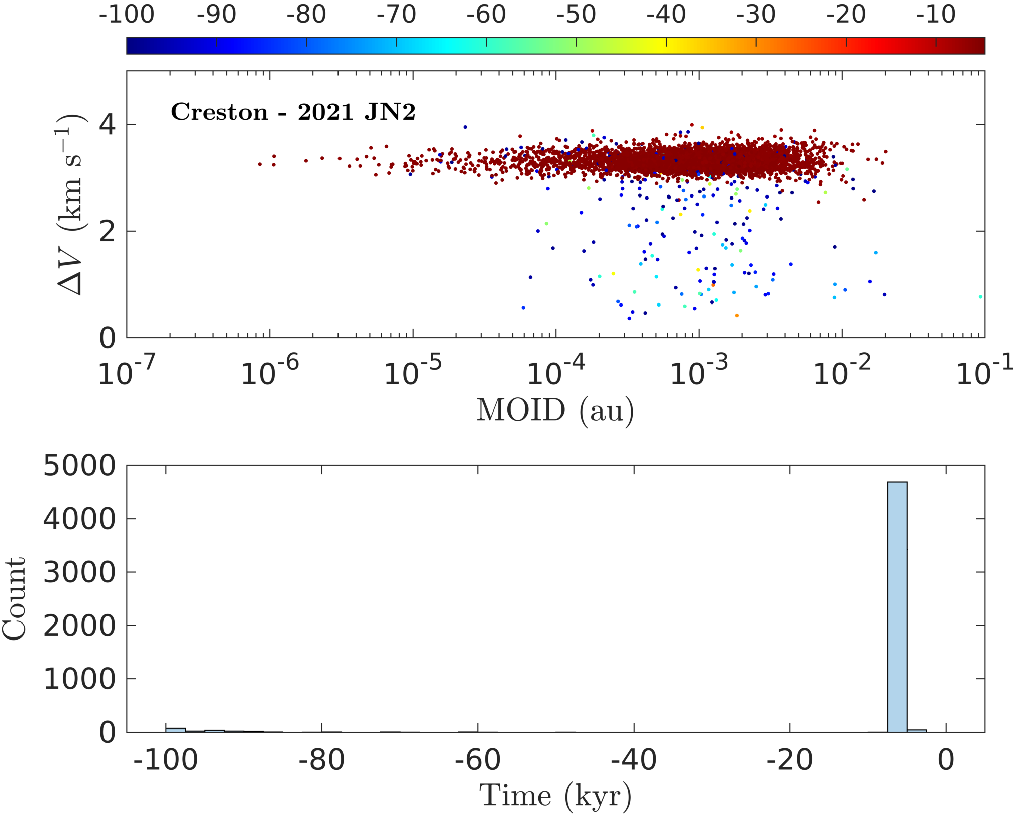}
    \vskip 10pt
    \caption{Distributions of the relative velocity vs. the MOID at the time $t_{\min}$ and distribution of $t_{\min}$, for the pairs Annama$-$2016 RX, Bunburra Rockhole$-$2021 FB, and Creston$-$2021 JN2.}
    \label{fig:Pairs_fig1}
\end{figure}

Compared with the cases examined in Sec.~\ref{ss:482488}, the evolution of the the nodal distances $R_a$ and $R_d$ as a function of $\omega$ covers a larger dense area (see Fig.~\ref{fig:5_met}, first row). This is caused mainly by a quite large $1\sigma$ uncertainty of 0.12 au in the semi-major axis of the orbit of the Annama meteorite (see Table~\ref{tab:meteorites}). The path followed by 2016 RX (green curve in Fig.~\ref{fig:5_met}) shows some deviations from the path followed by the meteorite clones, especially near $\omega = 180^\circ$ for the ascending node and near $\omega=0^\circ$ for the descending node. Moreover, a chaotic behaviour of the nominal orbit of 2016RX happening after several thousands of years of evolution can be seen.  
The first panel of Fig.~\ref{fig:Pairs_fig1} shows the distribution of the MOID and the relative velocity $\Delta V$ at the minima of the distance $d$ of Eq.~\eqref{eq:distAlbino}. Two separated concentrations of points stand out in the distribution: one at $\Delta V \sim 4$ km s$^{-1}$ and another one at $\Delta V < 4$ km s$^{-1}$, both spanning values of the MOID from $\sim$0.1 au down to $10^{-4}$ au. The minima at higher relative velocity are realized by about half of the clones within the first 2.5 ky of dynamical evolution, before strong chaotic dynamical effects are able to significantly change the secular evolution. On the other hand, the minima with smaller relative velocity appear to be concentrated at about $-35$ kyr, and they show a larger dispersion in $\Delta V$.
The two peaks in the time $t_{\min}$ suggest two possible separation epochs, and the values of the $\Delta V$ appear to be compatible with a collisional origin. 

\subsection{Bunburra Rockhole$-$2021 FB}
The evolution in the planes $(R_a, \omega)$ and $(R_d, \omega)$ are shown in the second row of Fig~\ref{fig:5_met}. The evolution of the possible parent body 2021 FB follows the densest area produced by the orbital clones of the Bunburra Rockhole meteorites, and the nodal distances do not show large variations during the secular evolution. Note also that the nominal orbital elements of 2021 FB (see Table~\ref{tab:neo_orbits}) are very close to that of the Bunburra Rockhole meteorite. 
The distribution of MOID and $\Delta V$ at the minimum of $d$, and the distribution of $t_{\min}$, are shown in Fig.~\ref{fig:Pairs_fig1}, third and fourth panels, respectively. A concentration of minima at MOID $\sim$0.05$-$0.07 au and $\Delta V \sim $3 km s$^{-1}$ can be seen in the distribution, and they are realized within the first 2.5 kyr of backward evolution by about $\sim$800 clones.  After this initial period of evolution, the minima appear to be almost uniformly spread in time, without showing any clear concentration. This is probably caused by strong chaotic effects dominating the secular dynamics, since these objects are placed inside the orbit of the Earth at relatively low inclination. Despite the concentration of minima of $d$ in the first 2.5 kyr, the relatively large values of the MOID suggest that this association is more likely to be spurious. In fact, Bunburra Rockhole is an Eucrite meteorite most likely coming from the innermost region of the main belt, and supposedly delivered through the $\nu_6$ secular resonance \citep{Bland2009}.

\subsection{Creston$-$2021 JN2}
\label{ss:creston}
The evolution of the candidate NEA parent body 2021 JN2 in the planes $(R_a, \omega)$ and $(R_d, \omega)$ shows a good agreement with the densest area traced by the orbital clones of the Creston meteorite (see Fig.~\ref{fig:5_met}, third row). As seen in the previous section, this is already an indication of a possible correlation between the two bodies. Figure~\ref{fig:Pairs_fig1} shows the distribution of the MOID and the relative velocity $\Delta V$ at the minima of the distance $d$, together with the times $t_{\min}$ at which the minima are realized. The outstanding feature of these two distribution is that almost all the minima are concentrated at $\Delta V \sim$ 3.5 km s$^{-1}$, with MOID spanning values between 0.01 au and $10^{-6}$ au. Moreover, all the minima are realized between $-5$ and $-7.5$ kys of dynamical evolution, that is likely small enough to ensure that the chaotic nature of the long term dynamics has not become prevalent. More extensive and accurate numerical simulations, that are beyond the scope of this work, would be needed to better assess the results presented here about this pair, possibly taking into account also the uncertainties in the orbit of the NEA parent body. The values of the relative velocity may be an indication that the Creston meteorite was formed by an impact on 2021JN2, however further observations and physical characterization would be needed to better explore this hypothesis.

\begin{figure}
    \centering
    \includegraphics[width=0.48\textwidth]{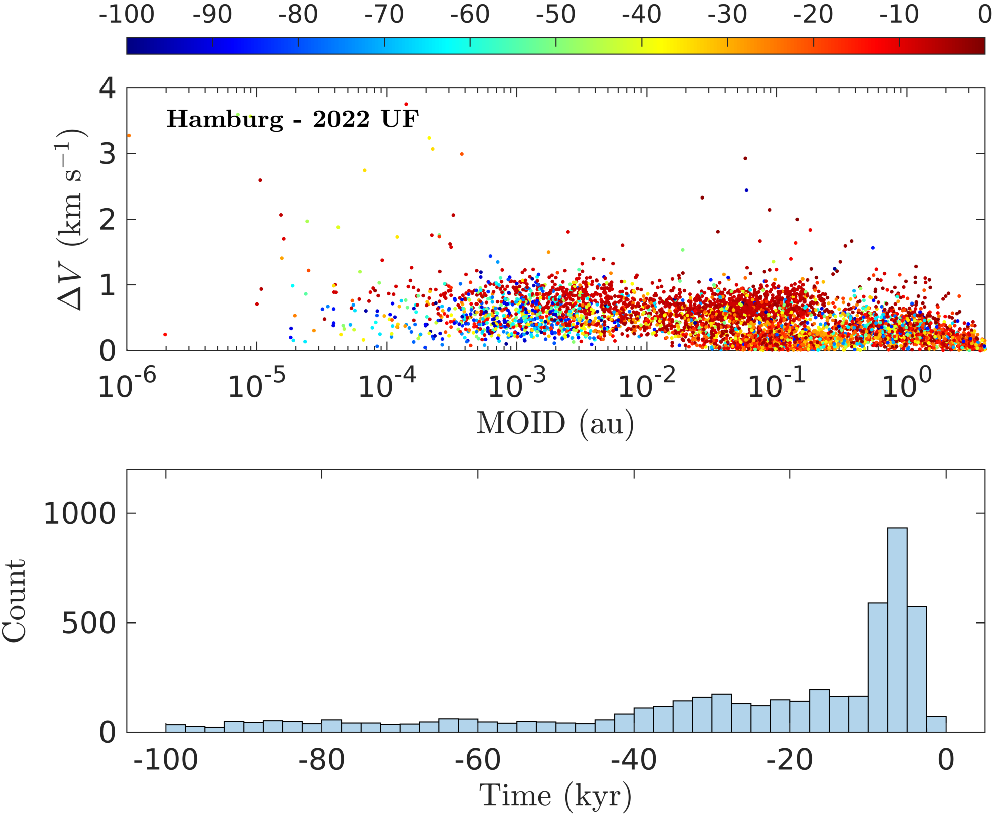}
    \vskip 10pt
    \includegraphics[width=0.48\textwidth]{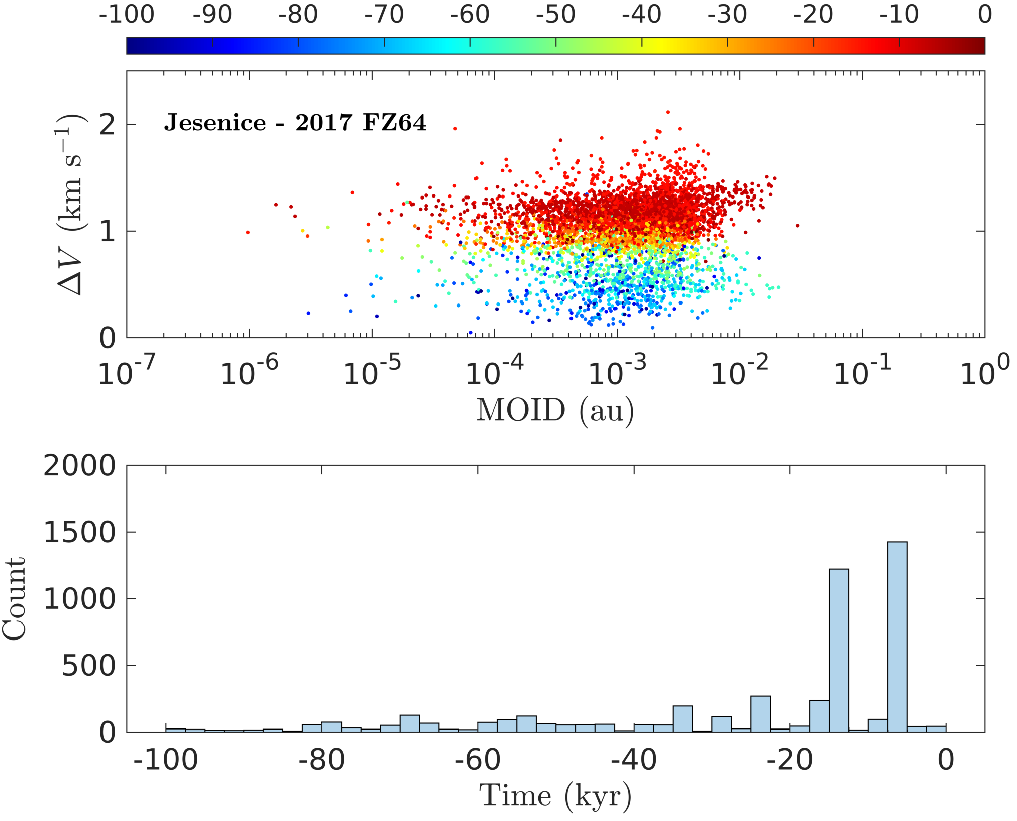}
    \vskip 10pt
    \includegraphics[width=0.48\textwidth]{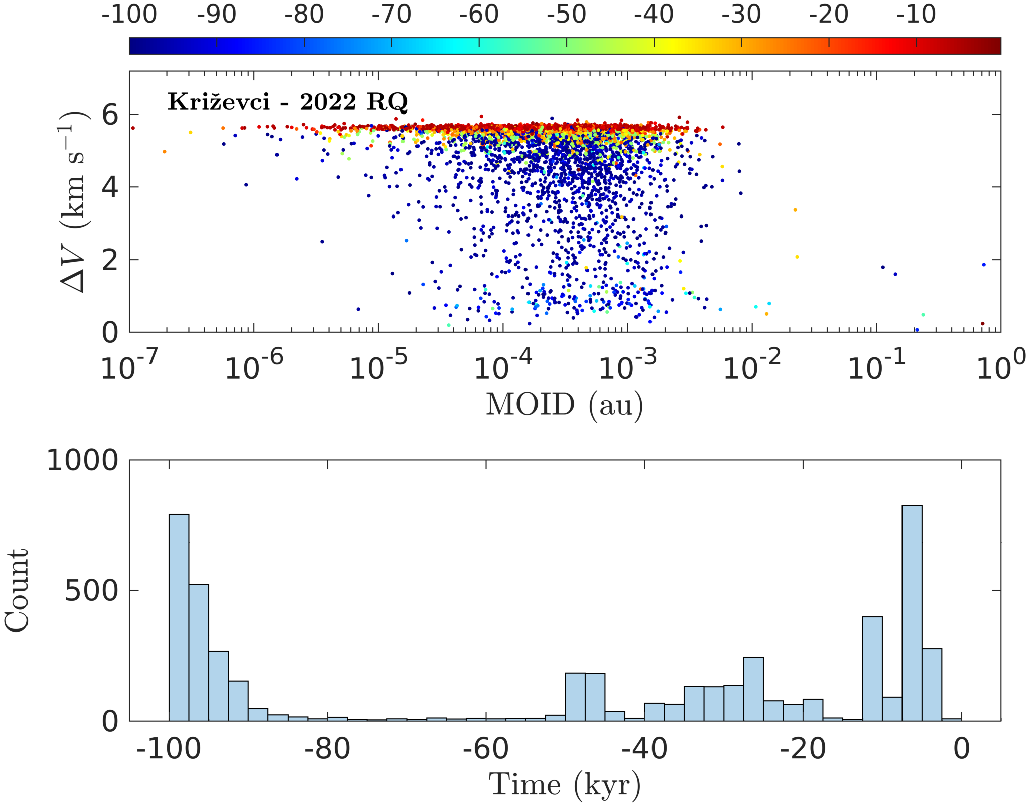}
    \vskip 10pt
    \caption{Distributions of the relative velocity vs. the MOID at the time $t_{\min}$ and distribution of $t_{\min}$, for the pairs Hamburg$-$2022 UF, Jesenice$-$2017 FZ64, and Križevci$-$2022 RQ.}
    \label{fig:Pairs_fig2}
\end{figure}

\subsection{Hamburg$-$2022UF, 2021 PZ1}
\label{ss:hamburg}
The fourth row of Fig.~\ref{fig:5_met} shows the density plot of the Hamburg meteorite clones in the planes $(\omega, R_a)$ and $(\omega, R_d)$, together with the path followed by the corresponding NEA parent body candidates. The primary 2022 UF is shown by green dots while the secondary 2021 PZ1 is represented by cyan dots. The candidate parent bodies and the meteorite clones have a similar qualitative evolution. However, we can see significant differences of more than 1 au between the ascending node $R_a$ of the two NEAs and the densest area of the meteorite clones distribution at $\omega = 180^\circ$, where $R_a$ is at the maximum. At $\omega=0^\circ$, where $R_a$ is at the minimum, the orbit of the NEAs is also a bit off of the densest part of the distribution, of about $\sim$0.5 au. Similar features can be seen also in the descending node $R_d$ at $\omega = 0^\circ$, where the nodal distance is at the maximum and the differences are of about $\sim$1 au, and at $\omega = 180^\circ$, where $R_d$ is at the minimum and the offset is of about $\sim$0.5 au.

In addition, we noted that 1591 clones of the Hamburg meteorite were expelled from the Solar System during the 100 kyr of backward evolution. The orbits of the clones are close to the 5:2 mean-motion resonance with Jupiter, that may be responsible of increasing the eccentricity sufficiently enough to put them on Jupiter crossing orbits and, in fact, the clones were ejected by effect of close encounters with Jupiter. We found that the distribution of the ejection times is fairly constant between $-10$ kyr and $-100$ kyr. This suggests that these orbits begin to be chaotic after a few kyr of dynamical evolution, making the attempt to identify their parent body of the Hamburg meteorite harder, unless the separation age happened only very recently. The distribution of the MOID and the relative velocity at the minima of $d$ for the pair Hamburg$-$2022 UF does show a concentration at $\Delta V < 1$ km s$^{-1}$ happened during the first $10$ kyr of dynamical evolution, before the start of the chaotic dynamical regime. The values of the MOID span a large interval of values, from $\sim$0.3 au down to $\sim 10^{-4}$ au, indicating that some of the meteorite clones arrive close enough to 2022 UF to suggest a possible common origin. On the contrary, the distribution for the pair Hamburg$-$2021 PZ1 does not show any concentration of points, and the times $t_{\min}$ are constantly distributed through the whole 100 kyr of dynamical evolution (see Fig.~\ref{fig:Pairs_Appendix}). Therefore, we believe this association to be spurious. As demonstrated by this case, and from others in the following sections, having a small value of $D_N$ is not enough to conclude the correlation, but further numerical simulations of the past evolution need to be performed. Note that an older separation event cannot be ruled out at this stage, however such simulations would require a more careful modeling of the dynamics that is out of the scope of this work. 

Apart from the chaotic nature of this specific case, the approach we used to find possible parent bodies may not give appropriate results in presence of objects that are placed deep inside a mean-motion resonance with Jupiter. Indeed, the method by \cite{2001Icar..152...58G} to compute the secular evolution of NEAs is valid under the assumption that no mean-motion resonances between the asteroid and a planet are present. High-order mean-motion resonances or mean-motion resonances with the inner planets may still not cause appreciable differences in the secular evolution. However, low-order mean-motion resonances with Jupiter are strong enough to significantly change the secular evolution, and another model for the secular propagation has to be used \citep{Fenucci2022}. Currently, the identification and the computation of proper elements for resonant NEAs is not automatized, and searching for possible parent bodies taking into account this additional dynamical aspect is beyond the purpose of this work.  

\subsection{Jesenice$-$2017 FZ64}
Figure~\ref{fig:5_met}, fifth row shows the evolution in the planes $(R_a, \omega)$ and $(R_d, \omega)$. As in the previous cases, the evolution of the candidate NEA parent body agrees with that of the meteorite clones. The distributions of the MOID and $\Delta V$ at the minima of $d$, as well as that of $t_{\min}$ (see Fig.~\ref{fig:Pairs_fig2}, third and fourth panels), show similar features to those of the pair Annama$-$2016 RX presented in Sec.~\ref{ss:annama}. The distribution of $t_{\min}$ has two sharp peaks at about $-10$ kyr and $-17$ kyr, with the first peak realized by about 1500 orbital clones and the second one by about 1200 orbital clones, indicating two possible separation epochs. The minima are concentrated at relative velocities of $\sim 1$ km s$^{-1}$, while the MOID spans values up to 0.02 au down to $10^{-5}$ au. Note that the values of relative velocity are still compatible with an impact origin.    

\begin{figure}
    \centering
    \includegraphics[width=0.48\textwidth]{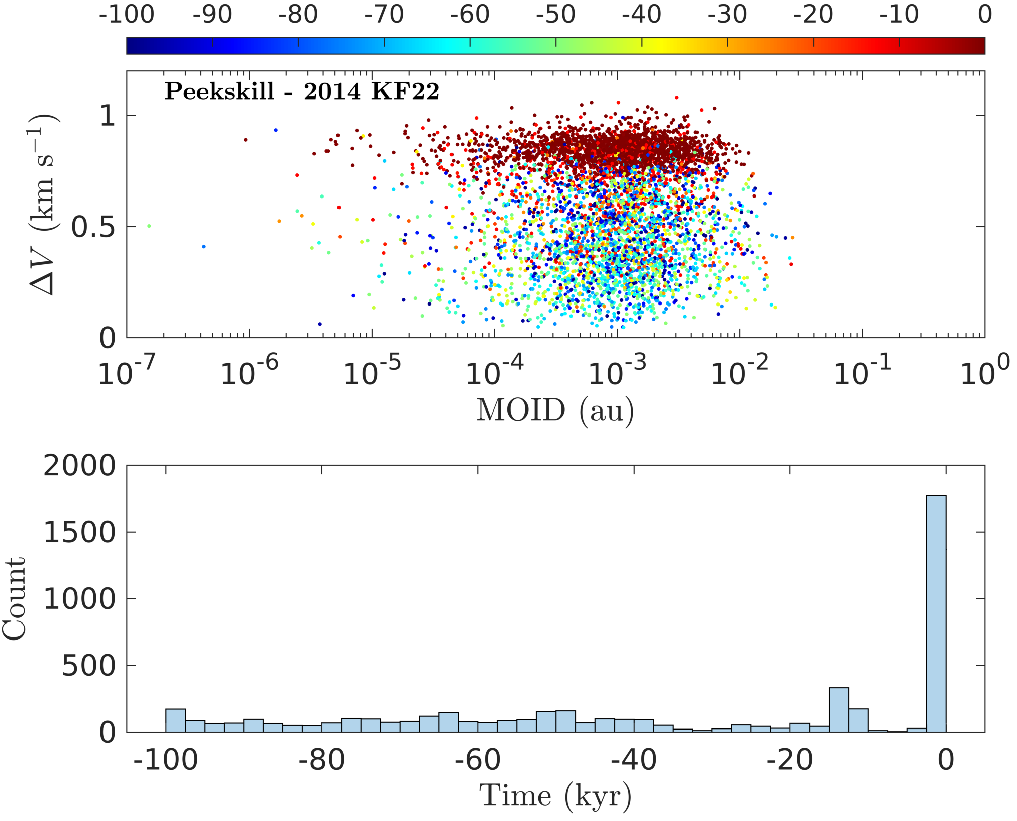}
    \vskip 10pt
    \includegraphics[width=0.48\textwidth]{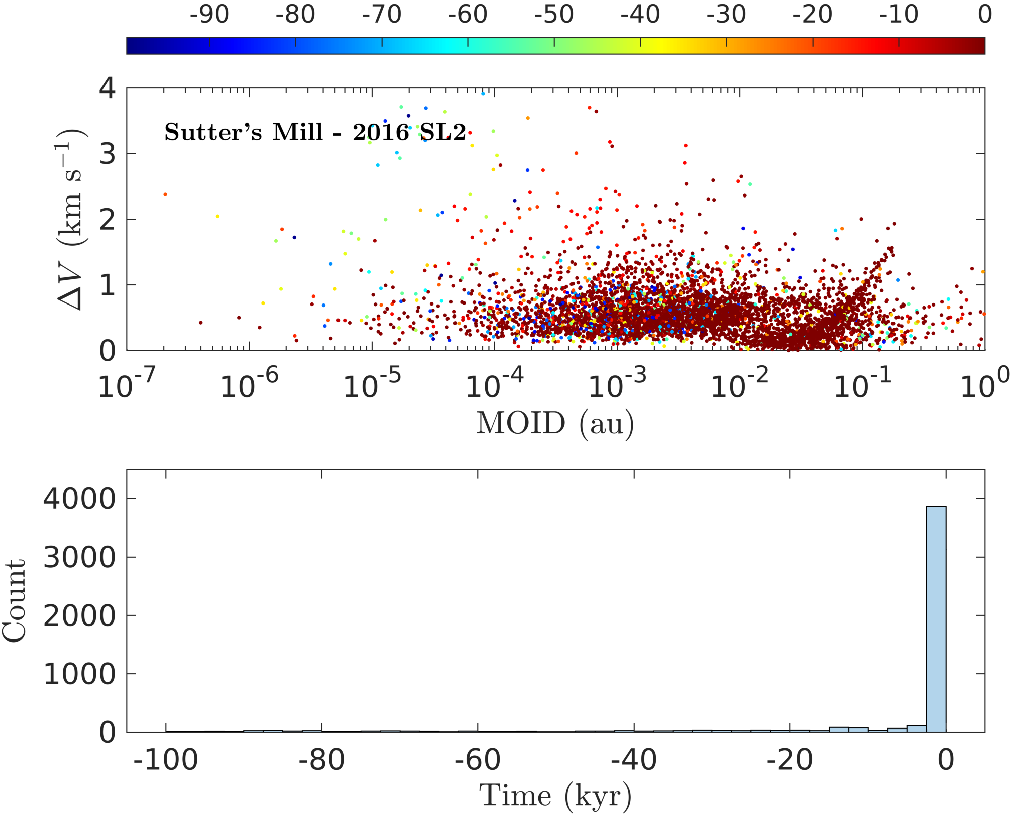}
    \vskip 10pt
    \includegraphics[width=0.48\textwidth]{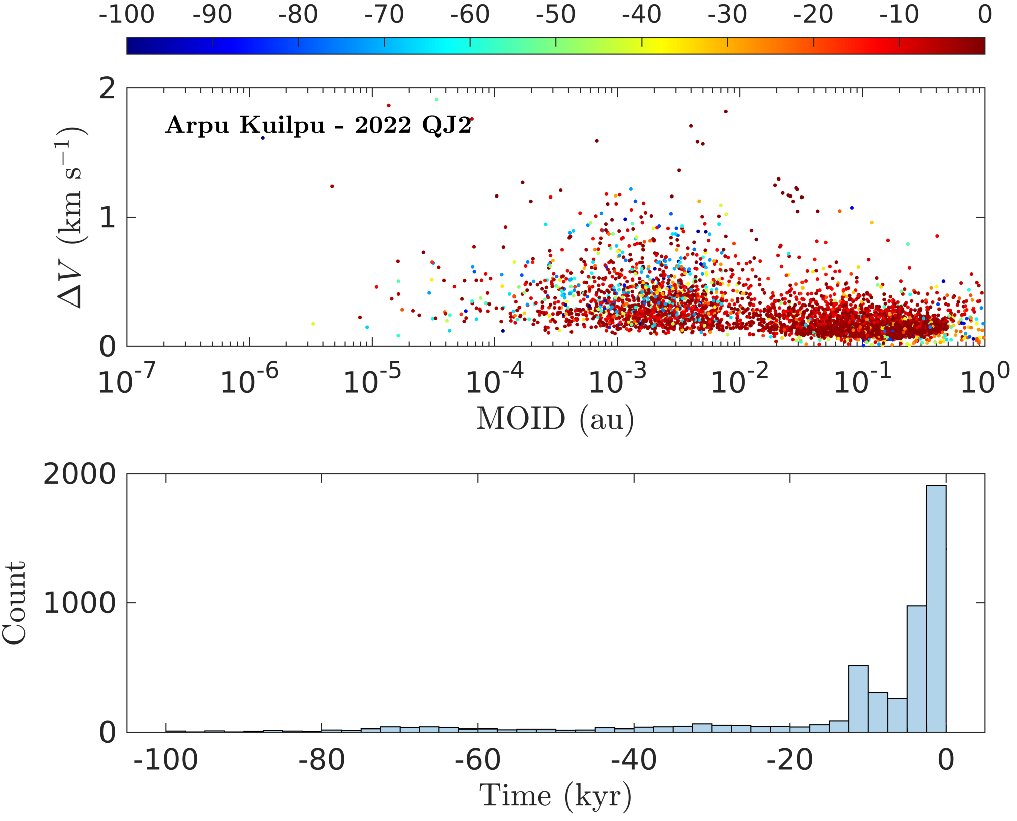}
    \vskip 10pt
    \caption{Distributions of the relative velocity vs. the MOID at the time $t_{\min}$ and distribution of $t_{\min}$, for the pairs Peekskill$-$2014 KF22, Sutter's Mill$-$2016 SL2, and Arpu Kuilpu$-$2022 QJ2.}
    \label{fig:Pairs_fig3}
\end{figure}

\subsection{Ko\v sice$-$2021 NV5, 2019 ST2}
The sixth row of Fig.~\ref{fig:5_met} shows the density plots in the planes $(\omega, R_a)$ and $(\omega, R_d)$, together with the path followed by the NEA parent body candidates 2021 NV5 and 2019 ST2 in green and cyan, respectively. This case is qualitatively similar to that of the pair Hamburg$-$2022 UF, 2021 PZ1 discussed in Sec.~\ref{ss:hamburg}, therefore we refer to the relative paragraph for the description of this aspect. 
In addition, 1180 orbital clones of the Košice meteorite were expelled from the Solar System, again probably by the effect of the 5:2 mean-motion resonance with Jupiter, with ejection times almost equally distributed between $-10$ kyr and $-100$ kyr. However, differently from the case of the Hamburg meteorite, we did not find any concentration of minima of $d$ before the beginning of the chaotic dynamics phase, for both the candidate parent bodies 2021 NV5 and 2019 ST2. For the distributions of the relative velocity vs. the MOID at the time $t_{\min}$ and distribution of $t_{\min}$, see Fig.~\ref{fig:Pairs_Appendix}. Therefore, we cannot give any strong evidence for a true correlation of these objects. According to \cite{Borovicka2013} the origin of the Košice meteoroid is in the central main asteroid belt near the 8:3 main motion resonance with Jupiter.

\begin{figure*}
    \centering
    \includegraphics[width=\textwidth]{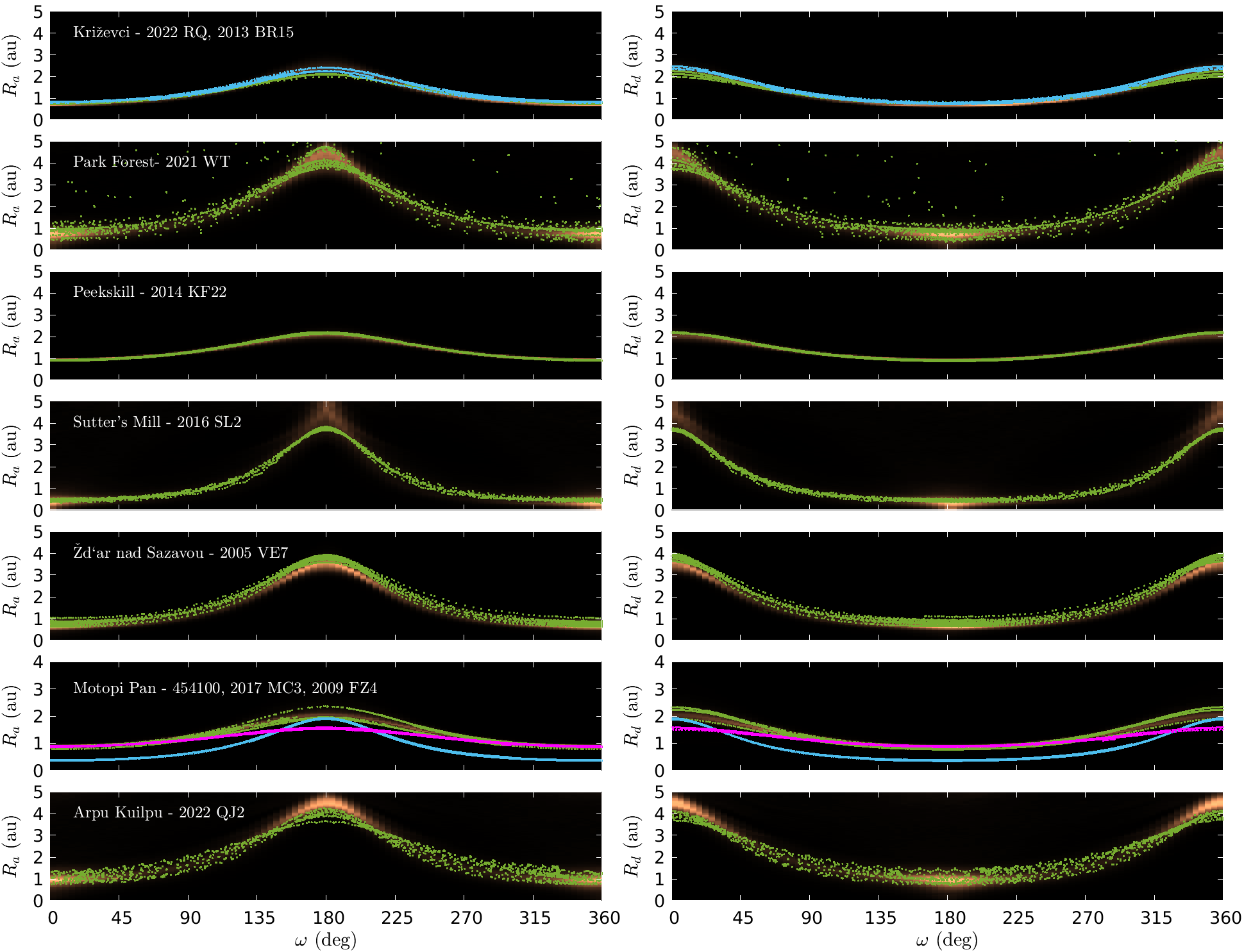}
    \caption{Same as Fig.~\ref{fig:Pribram_Neusch_482488} for the pairs Košice$-$2021 NV5, 2019 ST2 (first row), Kri\v zevci$-$2022 RQ, 2013 BR15 (second row), Park Forest$-$2021 WT (third row), Peekskill$-$2014 FK22 (fourth row), Sutter's Mill$-$2016 SL2 (fifth row), and \v Zd'ár nad Sázavou$-$2005 VE7 (sixth row). The green color represents the evolution of the primary associated NEA, and the cyan color the evolution of the secondary associated NEA, when present.}
    \label{fig:5_met2}
\end{figure*}

\subsection{Kri\v zevci$-$2022 RQ, 2013 BR15}		 
 The secular evolution of 2022 RQ and 2013 BR15 in the planes $(R_a, \omega)$ and $(R_d, \omega)$ are depicted in the first row of Fig.~\ref{fig:5_met2}, with green and cyan colors respectively. The evolution of both these NEAs follow the densest area covered by the orbital clones of the Križevci meteorite. Figure~\ref{fig:Pairs_fig2} shows the distribution of the MOID and $\Delta V$ at the minima of $d$ for the pair Križevci$-$2022 RQ, and a concentration of about 1000 points at about $-7.5$ kyr, gathered at $\Delta V \sim 6$ km s$^{-1}$, can be recognized. Other concentrations of minima of $d$ can be seen at times near $-100$ kyr, that can be attributed to the scattering of orbital elements due to chaos, and therefore probably not related to a true separation event. The relative velocity of almost $6$ km s$^{-1}$ may be an indication that the separation occurred because of an impact with a comparatively high relative velocity, although the possibility of a random association between the meteorite and NEA cannot be totally ruled out. 

The same analysis performed for the pair Križevci$-$2013 BR15 showed that the minima of $d$ are all concentrated at $-100$ kyr, except for about 300 clones that had a minima at about $-17$ kyr with relative velocity of $6$ km s$^{-1}$ (see Fig.~\ref{fig:Pairs_Appendix}). While this last mentioned concentration may be an indication of a separation induced by an impact, the small number statistics makes the conclusion of a true correlation between these bodies questionable, and therefore we tend to favor the random association hypothesis. 

\subsection{Park Forest$-$2021 WT}
The secular evolution of 2021 WT follows exactly the same evolution of the clones of the Park Forest meteorite in the planes $(R_a, \omega)$ and $(R_d, \omega)$, as shown in Fig.~\ref{fig:5_met2}, second row. However, the distribution of the minima of $d$ in the plane $(\text{MOID}, \Delta V)$ did not show any concentration of points, and the distribution of the times $t_{\min}$ did not have any concentration of times (see Fig.~\ref{fig:Pairs_Appendix}), suggesting that this was a random spurious association. Note that the Park Forest meteorite falls in a region of the plane $(\theta, \phi)$ that is quite filled by known NEAs (see Fig.~\ref{fig:meteorites_radiants}), providing an additional evidence of the spurious association hypothesis.

\subsection{Peekskill$-$2014 FK22}
The evolution of the nodal distances as a function of $\omega$ for the pair Peekskill$-$2014 KF22 are shown in the third row of Fig.~\ref{fig:5_met2}. Due to to the small uncertainties in the orbit of the Peekskill meteorite, the densest area covered by the meteorite clones results to be very narrow. The path of the candidate parent body 2014 FK22 follows almost exactly the same corridor indicated by the meteorite clones, suggesting already a possible common origin. 

Figure~\ref{fig:Pairs_fig3}, first panel, shows the distribution of the MOID and $\Delta V$ at the MOID for the minima of the distance $d$. The relative velocities are all smaller than 1 km s$^{-1}$, and a concentration of points appears at values of the MOID between $10^{-2}$ and $10^{-5}$ au, with relative velocities at around 0.8 km s$^{-1}$. The distribution of $t_{\min}$ is reported in the second panel of Fig.~\ref{fig:Pairs_fig3} and it shows a peak in the first 2.5 kyr of backward evolution, suggesting a really recent separation.

%\begin{figure}
%    \centering
%    \includegraphics[width=0.48\textwidth]{figures/peekskill_mindistalbino_new.eps}
%    \vskip 10pt
%    \includegraphics[width=0.48\textwidth]{figures/34_Sutter_Mill.eps}
%    \caption{Distributions of the value of the MOID vs. the relative velocity at the MOID at the time $t_{\min}$, and distribution of $t_{\min}$, for the pairs Peekskill$-$2014FK22 and Sutter's Mill$-$2016SL2. }
%    \label{fig:peekskill-suttersmill}
%\end{figure}

\subsection{Sutter's Mill$-$2016 SL2}
The evolution of the Sutter's Mill meteorite clones show some dispersion in the planes $(R_a, \omega)$ and $(R_d, \omega)$ (see Fig.~\ref{fig:5_met2}, fourth row), probably due to the relatively large uncertainties on the orbital elements (see Table~\ref{tab:meteorites}). Despite this, the evolution of 2016 SL2 is still compatible with that of the meteorite clones. Moreover, the distribution of $t_{\min}$ (see Fig.~\ref{fig:Pairs_fig3}) shows a concentration of 4000 orbital clones in the first 2.5 kys of backward evolution, thus likely before the beginning of the chaotic phase of the long term dynamics. The values of the MOID span an interval between 10$^{-5}$ au up to 0.1 au, with relative velocities concentrated mostly at $\Delta V < 1$ km s$^{-1}$. These values of $\Delta V$ are compatible with either a collisional origin or an ejection by YORP spin-up, however it is not possible to conclude a clear preference among the two possibilities due to the lack of data about the lightcurve of the candidate NEA parent body. As for the case of Creston$-$2021 JN2 presented in Sec.~\ref{ss:creston}, further more accurate numerical simulations should be performed to deeply assess the relation between these two objects.

\subsection{\v Zd'ár nad Sázavou$-$2005 VE7}
As in almost all the previous cases, the path of the NEA candidate parent body follows the densest area of the meteorite clones in the planes $(R_a, \omega)$ and $(R_d, \omega)$, see Fig.~\ref{fig:5_met2} fifth row. However, the distribution of MOID and $\Delta V$ at the minima of the distance $d$ do not show any clear accumulation of points at low velocity. The histograms of $t_{\min}$ (not shown in the paper) look almost equi-distributed throughout the first 80 kyr of backward dynamical evolution, showing a peak only near epoch $-100$ kyr. In this case, the concentration at such large epoch in the past is mostly due to the scattered dynamics induced by chaos, rather than from a true separation event, and therefore we cannot conclude a possible correlation for this pair. For the distributions of the relative velocity vs. the MOID at the time $t_{\min}$ and distribution of $t_{\min}$, see Fig.~\ref{fig:Pairs_Appendix}.

\subsection{Motopi Pan$-$454100, 2017 MC3, 2009 FZ4}
Motopi Pan is the only meteorite for which we found three candidate parent bodies. Figure~\ref{fig:5_met2}, sixth row, shows the evolution of the nodal distances $R_a$ and $R_d$ as a function of $\omega$ of the meteorite clones. The evolution of the same quantities of the candidate parent bodies 454100, 2017 MC3, and 2009 FZ4 are superimposed with green, cyan, and magenta dots, respectively. The best match is achieved by 454100, that follows very precisely the densest area given by the meteorite clones. The asteroid 2009 FZ4 also shows a good agreement in the evolution of the nodal distances, except near $\omega = 180^\circ$ for the ascending nodal distance and near $\omega = 0^\circ$ for the descending nodal distance. On the other hand, the evolution of 2017 MC results to be qualitatively different, and the crossing configurations with the Earth orbit happening at $R_a, R_d =1 $ au happen at values of $\omega$ much different from the other two cases.

\begin{figure}
    \centering
    \includegraphics[width=0.48\textwidth]{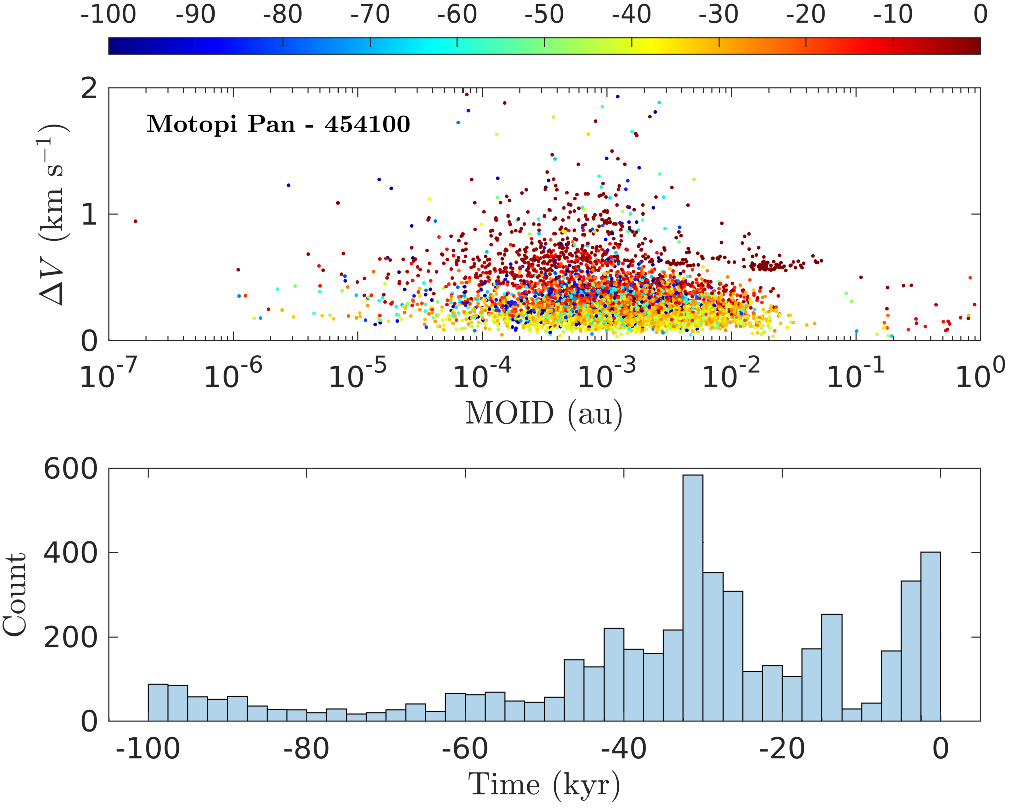}
    \vskip 10pt
    \includegraphics[width=0.48\textwidth]{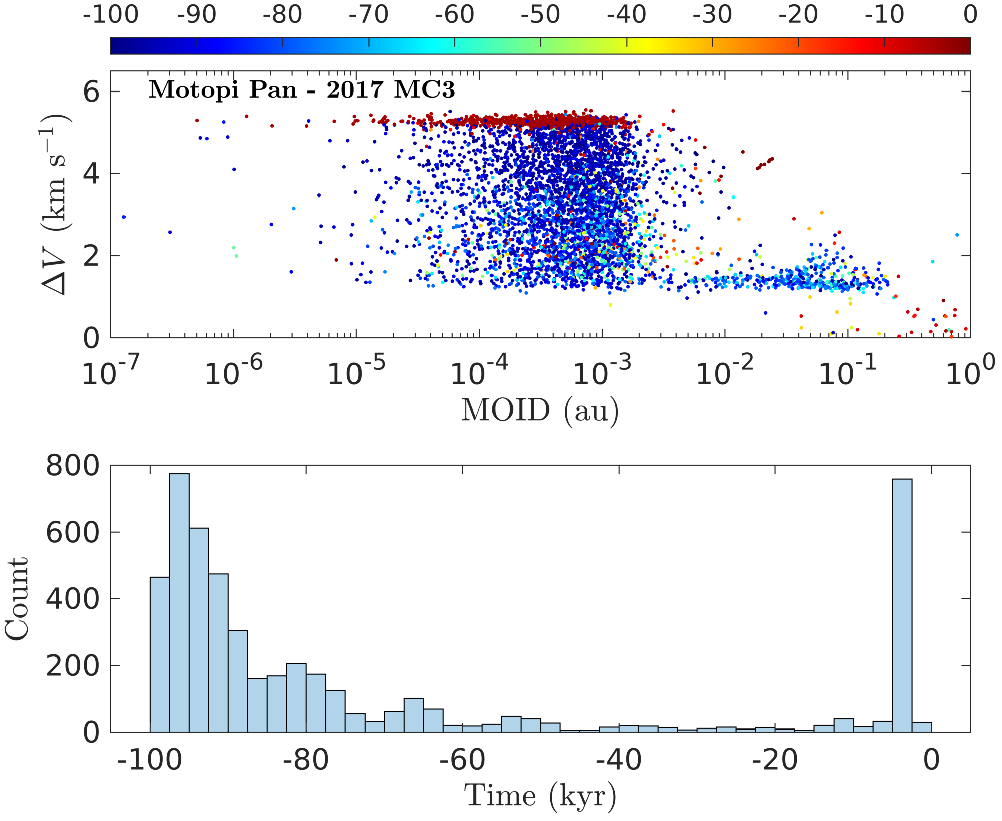}
    \vskip 10pt
    \includegraphics[width=0.48\textwidth]{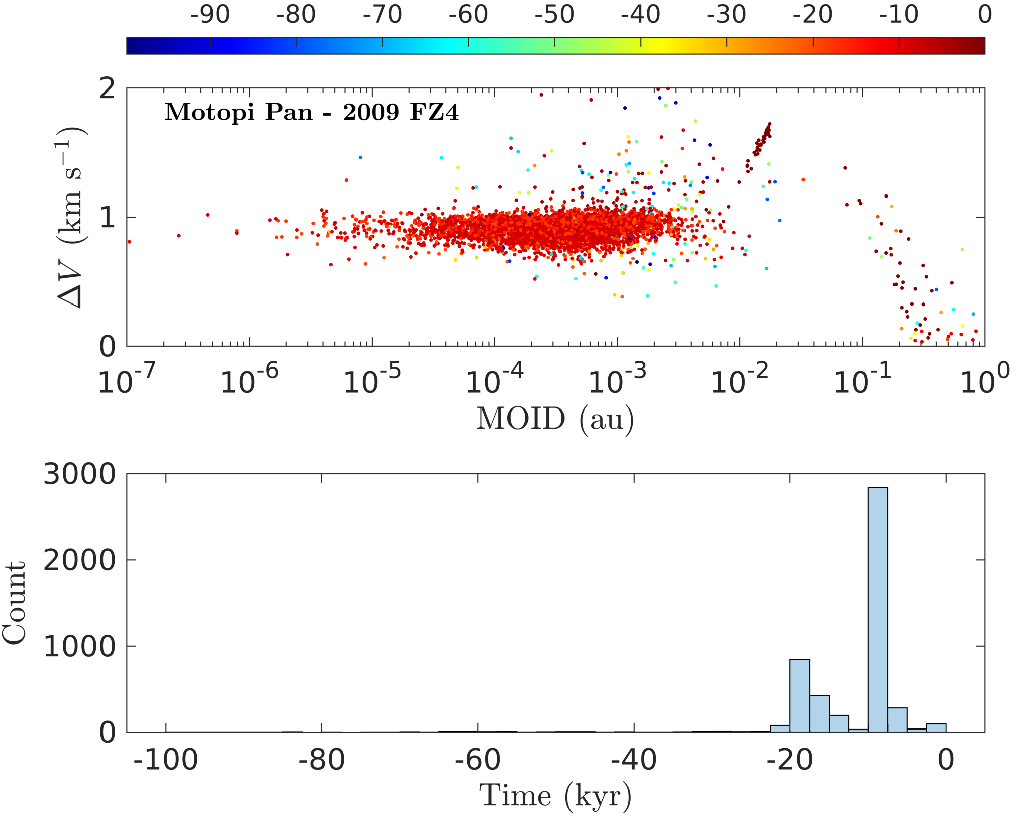}
    \vskip 10pt
    \caption{Distributions of MOID vs. relative velocity at the MOID at time $t_{\min}$, and distribution of $t_{\min}$, for the pairs Motopi Pan$-$454100, Motopi Pan$-$2017 MC3, and Motopi Pan$-$2009 FZ4.}
    \label{fig:Pairs_fig4}
\end{figure}

Figure~\ref{fig:Pairs_fig4} shows the evolution of MOID and $\Delta V$ at the minima of $d$, and the histograms of the time $t_{\min}$ for the three pairs. The minima for the couple Motopi Pan$-$454100 are all notably constrained at low relative velocities smaller than 1 km s$^{-1}$, while values of the MOID are rarely larger than 0.01 au. This suggests that the meteoroid that originated the Motopi Pan meteorite is very likely related to 454100. This result is consistent with what suggested by \cite{deLaFuente2019}. The small relative velocities are also compatible with a rotational fission origin, that could be in turn suggested by rotation period estimates obtained from photometry (see Sec.~\ref{s:discussion}). On the other hand, extrapolating a clear separation age from the histogram of $t_{\min}$ is not an easy task. Two concentrations appear between 0 and $-$5 kyr and at around $-30$ kyr of evolution, however their statistics is poor. Overall, only a few percent of minima are achieved after roughly $50$ kyr of dynamical evolution, and therefore we could only conclude that the separation event most likely happened in the last 50 kyr. A more accurate and thorough dynamical analysis, including also orbital clones for the candidate NEA parent body, would be required to better determine the separation age.

The pair Motopi Pan$-$2017 MC3 shows a concentration of minima at relative velocity $\Delta V \sim 5$ km s$^{-1}$, happening between $-5$ and $-7.5$ kyr. Other concentrations with low MOID values are achieved after 80 kyr of backward dynamical evolution, however they are vastly spread in relative velocity, spanning values of $\Delta V$ from $\sim 1.5$ km s$^{-1}$ up to $\sim 5$ km s$^{-1}$. Therefore, while simulations show that the two objects may be somewhat related, the separation event appears to be different from that that originated the pair Motopi Pan$-$2017 MC3.

Finally, the pair Motopi Pan$-$2009 FZ4 is qualitatively similar to Creston$-$2021 JN2. A large concentration of minima of $d$, gathered around $\Delta V \sim 1$ km s$^{-1}$ and at MOID values smaller than 0.005 au, appears at $-10$ kyr of dynamical evolution. This also suggest a possible correlation between the two objects, compatible with both a collision event or a rotational fission. 

Another plausible hypothesis is that the meteoroid that originated Motopi Pan, 2017 MC3, and 2009 FZ4 all come from 454100, which is the largest asteroid among them, and they all originated during the same event. Additional more accurate numerical simulations would be needed to explore this scenario.

\subsection{Arpu Kuilpu$-$2022 QJ2}
Figure~\ref{fig:5_met2}, seventh row, shows the evolution of the orbital clones of the meteoroid associated to the Arpu Kuilpu meteorite in the planes $(R_a, \omega)$ and $(R_d, \omega)$. The evolution of the candidate parent body 2022 QJ2 is depicted by green dots. As in most of the previous cases, the two paths are in good agreement. Figure~\ref{fig:Pairs_fig3} shows the distribution of MOID and $\Delta V$ at the minima of $d$. All the minima are attained at low relative velocities, however the MOID spans values as low as $\sim 10^{-4}$ au up to $\sim 1$ au, which seem to be too large to suggest a possible correlation. Note also that these orbits are deeply placed in the 3:1 mean-motion resonance with Jupiter \citep[see e.g.][]{Fenucci2023}, and therefore concluding an association based on the properties of the secular evolution is harder, as pointed out above. We consider therefore this association as spurious.
In addition, we found that during the backward evolution 1254 meteorite clones were expelled from the Solar System, by effect of deep close encounters with Jupiter. This suggests that the orbit of the meteoroid that originated the Arpu Kuilpu meteorite can be traced back only for a limited amount of time due to chaos produced by close approaches. We found that the expulsion times all happen after 10 kyr of backward dynamical evolution.

\section{Discussion}
\label{s:discussion}
\begin{figure}
\centering
\includegraphics[width=1.0\hsize]{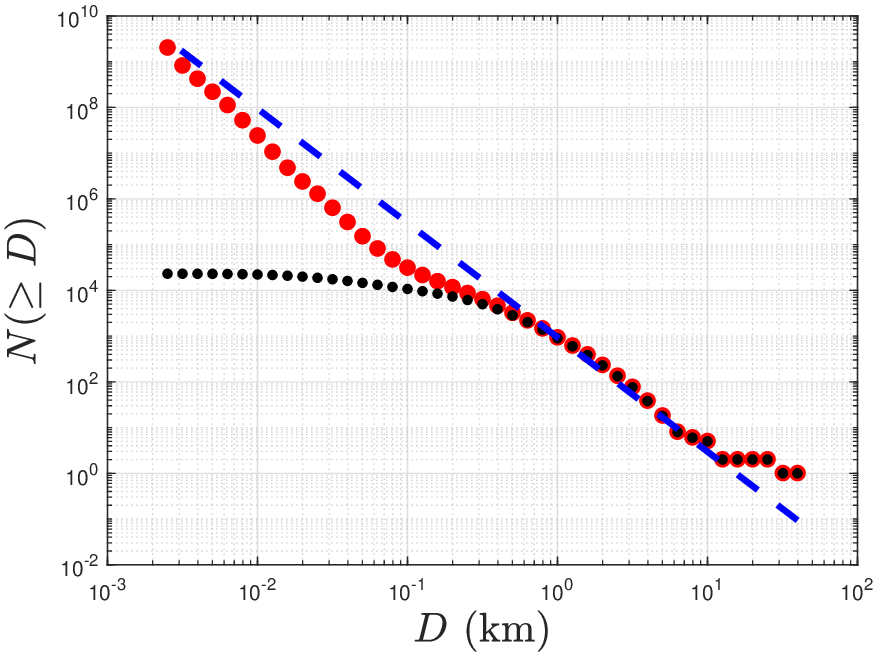}
\caption{A log-log plot of the NEAs cumulative diameter distribution according to \protect\citet{Harris2021} model (red dots) and the observed ones (black dots). The dashed blue line is the simple power law $N(\geq D)= 940\times D^{-2.5}$. This exponential law is what one would expect in the case of a population of bodies that has undergone a collisional evolution in which the destructive process depends only by the collision speed and the size ratio between the colliding bodies \protect\citep{Pater2010}. The constant 940 normalizes it to the \protect\citet{Harris2021} model for asteroids 1 km in diameter or larger. }
\label{fig:cumulative}
\end{figure}

Our results from the previous sections are summarized in Table~\ref{tab:Separation}. On a total of 20 possible meteorite$-$NEA pairs, in 8 cases it was not possible to say whether the association is real or not due to the orbital chaos. In the remaining 12 meteorites cases a physical association appear possible, with a range of minimum speed differences from 0.5 to 6 km s$^{-1}$: the lowest speed is for the Sutter's Mill$-$2016SL2 pair, while the highest is for Križevci$-$2022RQ. Anyway the  characteristic orbital speed difference is about 1 km s$^{-1}$, a value that appear compatible with an impact event.  \\
The ages of the collisions appear to be on the order of tens of thousands of years, a much lower value than the dynamic life time of a NEA of about 10 Myrs \citet{Gladman1997}, therefore the separation must have occurred when the progenitor was already part of the NEAs population. The case of Pribram and Neuschwanstein is peculiar because the age of separation from the asteroid 482488 appear similar, 20$-$30 kyr, so they were probably born in the same impact event. In Section~\ref{ss:482488} we already noted that, even if the cosmic-ray exposure time of the two meteorites are different (12 Myr Pribram, 48 Myr Neuschwanstein), this does not prevent them from having a common origin. Most likely 482488, considering the diameter, is a rubble pile asteroid as Bennu, made up of blocks with very different and complex exposure histories. Also very interesting is the case of Motopi Pan which appears to be associated with three asteroids, even if the sequence of separation between these bodies is not clear. Also for Motopi Pan the cosmic-ray exposure time is high and estimated to be $19.2\pm 2.4$ Myr \citep{Jenniskens2021}. If we suppose that Motopi Pan, 2017 MC3 and 2009 FZ4 were born from the same collision event between an unknown asteroid and 454100 which, being the largest of the three must be considered the main asteroid, then it is likely that the separation age is between $-$2 kyr and $-$10 kyr and that there are other small meteoroids on similar orbits that could collide with the Earth in the future. Effectively, according to \cite{deLaFuente2019}, it is possible that 2018 LA is associated with 454100 and the $\chi-$Scorpiids meteor shower.\\
Now the question that needs to be answered is: this typical separation time of tens of thousands of years, is it consistent with impact events? To establish if the typical separation time value found from the orbital analysis of meteorites and NEAs appears reasonable we use a simple collision model to estimate the expected collision frequency for a NEA of about 0.1 km in diameter (the typical size of Table~\ref{tab:ProgenitorCandidates}) which undergoes a collision with micro-NEAs of the order of 2 m in diameter. The starting point is the NEAs population model of \citet{Harris2021}, see Fig.~\ref{fig:cumulative} for a plot. 

\noindent From the cumulative distribution of this model we have $N_1\left(\geq D_1 = 0.1 ~\hbox{km}\right) \approx 3.1\times 10^4$ (type 1 NEAs) and $N_2\left(\geq D_2 = 0.0025 ~\hbox{km}\right) \approx 2.05\times 10^{9}$ (type 2 NEAs). To estimate the collision frequency of type 1 NEAs immersed in a population of type 2 NEAs, we assume a simple particle-in-a-box model with type 1 NEAs fixed, and type 2 NEAs moving randomly with relative velocity $\Delta V \approx 15$ km s$^{-1}$ within an interaction volume $V_{\text{int}}$ \citep{Wetherill1967, Farinella1992}:

\begin{equation}
f=\frac{\pi N_1 N_2 \left(D_1+D_2\right)^2 \Delta V}{4 V_{\text{int}}}.
\label{eq:nea_frequency}
\end{equation}

\noindent In Eq. (\ref{eq:nea_frequency}) $N_1$ is the bodies number of the target population, $N_2$ the bullets population while $D_1$ and $D_2$ are their respective diameters. The model represented by Eq. (\ref{eq:nea_frequency}) has some limits: there is not a well-defined volume within the NEAs can move, and the relative motion is not completely random as occurs in a gas, however the model can be adequate for estimates of orders of magnitude. A first estimate of the interaction volume for NEAs can be obtained by considering the explosions in the Earth's atmosphere reported in the CNEOS database\footnote{\url{https://cneos.jpl.nasa.gov/fireballs/}}. \\
As of Jan 31, 2023, there are 945 observed explosions in the atmosphere since April 15, 1988. On average, one atmospheric explosion every 13.445 days, with an estimate total impact energies ranging from 0.1 to 100 kt (1 kt = $4.184\times 10^{12}$ J), with a median value of about 0.2 kt. Most of these bodies are of asteroidal origin, therefore with typical atmospheric speeds of 10$-$20 km s$^{-1}$. If we assume an average value of about 15 km s$^{-1}$ and a bulk density of about $3500$ kg m$^{-3}$, the energy range 0.1$-$100 kt corresponds to a range of about 1$-$10 m in diameter, with a median value of about 2 m. As a typical diameter value for this population of very small NEAs we assume this value. \\
From the estimate number of this population given by \citet{Harris2021} and assuming that the observed collision frequency $f\approx 8.61\times 10^{-7} ~\hbox{s}^{-1}$ was not biased, reversing Eq. (\ref{eq:nea_frequency}) we can obtain $V_{\text{int}}$ using for $D_1$ the gravitational diameter of the Earth ($D_1 = 2 \times 7950 = 15~900 $ km) with $\Delta V\approx 15$ km s$^{-1}$. Considering that the Earth is only one, in this case $N_1 = 1$. Doing so we find $V_{\text{int}}\sim 7\times 10^{24} \approx 10^{25} ~\hbox{km}^3$, equivalent to a cube with a side of about 1.4 au. \\
Now with our estimate of the $V_{\text{int}}$ value, we can estimate the collision frequency of the NEAs with $D_1\geq$ 0.1 km ($N_1 \approx 3.1\times 10^4$) with the whole NEAs population with $D_2\geq$ 2 m ($N_2 \approx 2\times 10^9$). We found $f\approx 1.1\times 10^{-12} ~\hbox{s}^{-1}$, so we can expect a collision roughly every 30~000 years. This rough estimate makes it reasonable to find a certain number of meteorite-NEA pairs not older than a few tens of thousands of years.\\ 
We made an alternative estimate of $V_{\text{int}}$ assuming it is a torus with a square section and using the median values of $a$, $e$ and $i$ for the population of observed NEAs. The median values of the typical orbit are $a\approx 1.7$ au, $e\approx 0.45$ and $i\approx 9^{\circ}$, so the internal radius of a square section torus is $R_1=a(1-e)\approx 0.93$ au while the external one $R_2=a(1+e)\approx 2.46$ au. The average height of the orbit above ecliptic is $h=(R_2\sin(i)+R_1\sin(i))/2\approx 0.26$ au and the volume of the square section torus is $V_{\text{int}}=2 \pi h (R_2-R_1)(R_2+R_1)\approx 10^{25} ~\hbox{km}^3$, in agreement with the previous value estimated from the collision frequency of the micro-NEAs.\\
Based on these results, it seems reasonable to assume that the very large population of micro-NEAs collide with small NEAs, generating a part of the meteorites found on Earth. Our starting sample consisted of 38 meteorites (Table~\ref{tab:meteorites_encounters}) but only 10 of them appear associated with a known NEA, which means that about $10/38 \sim 25\%$ of the meteorites probably originate directly in the NEAs population, mainly from little collision events in the inner Solar System, rather than from asteroid collisions in the main belt. In summary, the collisions that occur in the main belt can also continue to occur between the NEAs, generating a minority part of the meteorites that we find on Earth.\\ 
In order to better characterize the possible connection from meteorites and NEAs, we invite the astronomers to observe at least the candidate parent objects with the lowest orbital uncertainty by taking advantage of the next favorable close encounter. Unfortunately most of the Table~\ref{tab:Separation} asteroids have very uncertain orbits and need to be recovered. Furthermore, given the small size, in general large telescopes are needed. The best candidates for observation are 2014K F22, which will be 0.02 au from Earth on May 25, 2025 with a magnitude of +21, and 454100 which will be 0.084 au on May 30, 2035. This latter NEA is quite large and will reach maximum brightness on May 20, with visual magnitude +17.2: photometric and spectroscopic observations could be carried out to determine both the rotation period and the spectral type of this very intriguing asteroid.\\
Finally, it is important to point out two features of the method used here. The first one is that the dynamics of objects in the NEA region is naturally subjected to chaos, with Lyapunov times often not larger than few tens of thousands years, and the method is expected to identify pairs with an age comparable to this timespan. Older pairs may be found only in really stable cases for which the Lyapunov time is much longer, but they would be very rare, considering also that the orbits of meteorites cross the orbit of the Earth by definition. \\
The second remark is that the association between meteorites and NEAs is naturally subjected to changes. While about 96\% of NEAs of 1 km in diameter are known, estimates suggest that only about 44\% of NEAs smaller than 140 m and the 5\% of NEAs smaller than 50 m have been discovered \citep{Harris2021}. Table~\ref{tab:ProgenitorCandidates} shows that asteroids as small as 10$-$20 m can still be dynamically associated to meteorites, thus more and more progenitor candidates can be found as the number of known NEAs increases. It is therefore important to attempt better dynamical associations as the catalogue of NEAs increases and their orbits improve in quality. 
 
\begin{table}
    \centering
    \caption{List of the candidate pairs meteorites-NEA with the possible separation age, minimum MOID and minimum $\Delta V$. Note that for all the meteorites with high orbital inclinations Pribram, Neuschwanstein, Bunburra, Jesenice and Annama it was possible to establish a separation event. I=Inconclusive results, S=spurious association.}
    \label{tab:Separation}
    \setlength\tabcolsep{2.5pt} % default value: 6pt
    \begin{tabular}{lcccc} 
    \hline
    Meteorite Name & NEA & Age (kyr) & MOID (au) & $\Delta V$ (km s$^{-1}$) \\
    \hline
    Pribram			      &	482488	 &  $-$30/$-$25      & $10^{-4}-10^{-3}$   & 1     \\
    Peekskill		      &	2014 KF22 &  $-$2.5           & $10^{-4}-10^{-2}$   & 0.8   \\
    Neuschwan.  	      &	482488	 &  $-$25/$-$20      & $10^{-4}-10^{-3}$   & 1     \\
    Park Forest			  & 2021 WT   &  S                & -                   & -     \\
    Bunburra Rockhole     & 2021 FB	 &  S                & -                   & -     \\
    Jesenice			  & 2017 FZ64 &  $-10$ or $-17$   & $10^{-5}-10^{-2}$   & 1     \\
    Ko\v sice			      &	2021 NV5	 &  I                & -                   & -     \\
                          & 2019 ST2  &  I                & -                   & -     \\
    Kri\v zevci		      &	2022 RQ	 &  $-$7.5           & $10^{-6}-10^{-3}$   & 6     \\
                          & 2013 BR15 &  S                & -                   & -     \\ 
    Sutter's Mill		  & 2016 SL2  &  $-2.5$           & $10^{-4}-10^{-1}$   & 0.5   \\            
    Annama			      &	2016 RX	 &  $-$2.5 or $-$35  & $10^{-3}-10^{-1}$   & 4 or 3\\
    \v Zd'ár nad S.          &	2005 VE7	 &  I                & -                   & - \\
    Creston			      &	2021 JN2	 &  $-$7.5/5         & $10^{-5}-10^{-2}$   & 3.5   \\
    Hamburg			      &	2022 UF	 &  $-$10/0          & $10^{-4}-10^{-1}$   & 1     \\
                          & 2021 PZ1  &  S                & -                   & -     \\
    Motopi Pan            & 454100   &  $-$2.5/$-$30     & $10^{-5}-10^{-2}$   & $<0.5$   \\
                          & 2017 MC3  &  $-$5             & $10^{-5}-10^{-2}$   &  5    \\
                          & 2009 FZ4  &  $-10$            & $10^{-6}-10^{-2}$   &  1    \\
    Arpu Kuilpu           & 2022 QJ2  &  S                &        -            &  -    \\
    \hline
    \end{tabular}
\end{table}

\section{Conclusions}
The aim of our work was try to determine possible parents NEAs for meteorites. We considered 38 meteorites whose heliocentric orbit was determined by triangulating their fireball during the fall and of which the coordinates of the true geocentric radiant and speed (which is not the entry speed into the atmosphere) are available. Starting from the encounter conditions of the Earth with the meteorites and the known NEAs population, $DN$ criterion based on geocentric quantities was computed with a certain limit threshold, finding 20 possible NEA$-$meteorite pairs. \\
The orbital evolution of these 20 pairs was studied starting from the time of the fall and going back in time. To account for the greater uncertainty of meteorite orbital elements with respect to NEAs, a population of 5000 clones was used for each meteorite. For 12 of these potential pairs, concentrations of distance minima were found in the phase space formed by the MOID of the orbits and by the relative speed at MOID. The minimum MOID between the orbits is around $10^{-4}$ au, while the relative velocity is about 1 km s$^{-1}$, with time scales of the order of tens of thousands of years for the NEA-meteorite separation time. The relative speed of about 1 km s$^{-1}$ suggests that collisions events separated the meteorite from the parent NEA. Other events that could generate separation, such as rotational instability, tidal destruction or thermal fracturing, would result in much lower relative speeds. Very interesting and unique are the Pribram and Neuschwanstein meteorites which, despite the diversity in cosmic-ray exposure time, probably separated from the same NEA with a rubble pile structure, (482488) 2012 SW20, about 20$-$30 kyr ago. The Motopi Pan meteorite is also very interesting, having three possible candidate NEA parent bodies: 454100, 2017 MC3, and 2009 FZ4. The NEA 454100 as progenitor of Motopi Pan was also found independently by \cite{deLaFuente2019}.\\
Based on a simple NEA-in-the-box model, it is reasonable to expect times of the order of a ten thousand years between collisions of NEAs of the order of 100 m in diameter and much smaller NEAs of the order of 2 m in diameter, in good agreement with what we found. So it seems reasonable to say that a part of the recovered meteorites, about 25\% of the starting meteorite sample according to our estimates, does not originate from collisions in the main belt, but from collisions events that occur directly in the population of small and very small NEAs in the inner Solar System. For this reason it is important to physically characterize the small asteroids probably associated with meteorites, even if they are objects which, even during close passages with the Earth, require large instruments to be observed. The possible meteorite-NEA pairs are also indicators of potential meteoroid showers whose radians $-$ if exist $-$ should be close to those of the meteorites. This work of matching between the radiants of meteorites and the radiants of already known meteor showers will be the subject of a forthcoming paper.

\section*{Acknowledgements}
The authors want to thank very much Giovanni Valsecchi (INAF-Institute of Space Astrophysics and Planetology), for the comments and suggestions that made the paper much better than the original. Many thanks to SpaceDyS and Fabrizio Bernardi for the use of the \texttt{OrbFit} code for NEAs-Earth encounter conditions computation. Finally our thanks go to the referee, whose suggestions make this paper much better than original.

%%%%%%%%%%%%%%%%%%%%%%%%%%%%%%%%%%%%%%%%%%%%%%%%%%
\section*{Data Availability}
The data underlying this article will be shared on reasonable request to the corresponding author.

%The inclusion of a Data Availability Statement is a requirement for articles published in MNRAS. Data Availability Statements provide a standardised format for readers to understand the availability of data underlying the research results described in the article. The statement may refer to original data generated in the course of the study or to third-party data analysed in the article. The statement should describe and provide means of access, where possible, by linking to the data or providing the required accession numbers for the relevant databases or DOIs.



%\bibliographystyle{mnras}

% This method is tedious and prone to error if you have lots of 
% references

\appendix

\section{Additional tables and figures}

\begin{table*}
	\centering
	\caption{Meteorites list with Earth encounter conditions $U$, $\theta$, $\phi$ and $\lambda$ from  true radian and geocentric velocity used to compute the distance $D_N$ with the NEAs whose orbit intersects Earth's orbit. Earth's encounter condition type: DePre = descending node before perihelion; DePost = descending node post perihelion; AsPost = ascending node post perihelion; AsPre = ascending node before perihelion.}
	\label{tab:meteorites_encounters}
	\begin{tabular}{llcccccccc} 
		\hline
		N & Name & Type & $U_x$ & $U_y$ & $U_z$ & $U$ & $\theta$ ($^\circ$) & $\phi$ ($^\circ$) & $\lambda$ ($^\circ$)\\
		\hline
01  &  Pribram                  & DePre	 &   -0.530	  &   0.136   &  -0.209   &  0.585   &   76.602  &  248.500	 &  197.827\\
02  &  Lost City                & DePost &    0.182	  &   0.223	  &  -0.380	  &   0.477	 &   62.088	 &  154.399	 &  104.486\\
03  &  Innisfree                & DePost &    0.013	  &   0.272	  &  -0.391	  &   0.477	 &   55.223	 &  178.072	 &  138.216\\
04  &  Benesov                  & DePre	 &   -0.330	  &   0.119	  &  -0.487	  &   0.601	 &   78.549	 &  214.129	 &  227.004\\
05  &  Peekskill                & AsPost &    0.305	  &   0.114	  &   0.095	  &   0.339	 &   70.326	 &   72.648	 &   16.836\\
06  &  Moravka                  & DePre	 &   -0.170	  &   0.016	  &  -0.636	  &   0.658	 &   88.619	 &  194.936	 &  226.304\\
07  &  Neuschwanstein           & DePre	 &   -0.527	  &   0.128	  &  -0.227	  &   0.588	 &   77.454	 &  246.648	 &  196.832\\
08  &  Park Forest              & DePre	 &   -0.511	  &   0.164	  &  -0.066	  &   0.541	 &   72.314	 &  262.609	 &  185.867\\
09  &  Villalbeto de la Peña    & DePost &	  0.513	  &   0.243	  &  -0.000	  &   0.567	 &   64.643	 &   90.016	 &  103.615\\
10  &  Bunburra Rockhole        & AsPost &	  0.140	  &  -0.113	  &   0.137	  &   0.226	 &  119.987	 &   45.620	 &  297.597\\
11  &  Maribo                   & DePre	 &   -0.850	  &  -0.072	  &   0.012	  &   0.853	 &   94.827	 &  270.821	 &  117.693\\
12  &  Jesenice                 & DePre	 &   -0.082	  &   0.177	  &  -0.199	  &   0.279	 &   50.485	 &  202.461	 &  199.232\\
13  &  Grimsby                  & DePost &	  0.163	  &   0.077	  &  -0.573	  &   0.601	 &   82.641	 &  164.141	 &    2.953\\
14  &  Košice                   & DePre	 &   -0.226	  &   0.258	  &  -0.044	  &   0.346	 &   41.726	 &  258.936	 &  160.085\\
15  &  Križevci                 & DePre	 &   -0.481	  &   0.061	  &  -0.012	  &   0.485	 &   82.767	 &  268.572   &	135.641\\
16  &  Sutter's Mill            & DePost &	  0.868	  &  -0.089	  &  -0.038	  &   0.873	 &   95.831	 &   92.528   &	212.704\\
17  &  Novato                   & AsPost &	  0.110	  &   0.223	  &   0.119	  &   0.276	 &   35.992	 &   42.966   &	 24.954\\
18  &  Chelyabinsk              & DePost &	  0.494	  &   0.073	  &  -0.095	  &   0.508	 &   81.697	 &  100.845   &	146.425\\
19  &  Annama                   & DePre	 &   -0.675	  &   0.011	  &  -0.255	  &   0.722	 &   89.111	 &  249.320   &	208.600\\
20  &  Žd'ár nad Sázavou        & DePre	 &   -0.618	  &   0.062	  &  -0.053	  &   0.623	 &   84.283	 &  265.137   &	 77.304\\
21  &  Porangaba                & DePost &	  0.317	  &   0.210	  &  -0.185	  &   0.423	 &   60.210	 &  120.304   &	108.927\\
22  &  Sariçiçek                & DePre	 &    0.000	  &   0.058	  &  -0.436	  &   0.440	 &   82.466	 &  179.943   &	339.830\\
23  &  Creston                  & AsPre	 &   -0.369	  &   0.026	  &   0.076	  &   0.378	 &   86.026	 &  281.624   &	 30.282\\
24  &  Murrili                  & AsPost &	  0.055	  &   0.261	  &   0.074	  &   0.277	 &   19.531	 &   36.839   &	 64.641\\
25  &  Ejby                     & DePre	 &   -0.167	  &   0.269	  &  -0.022	  &   0.317	 &   32.014	 &  262.643   &	137.289\\
26  &  Dishchii'bikoh           & DePost &	  0.178	  &  -0.031	  &  -0.371	  &   0.413	 &   94.267	 &  154.373   &	252.112\\
27  &  Dingle Dell              & DePre	 &   -0.270	  &   0.210	  &  -0.086	  &   0.353	 &   53.409	 &  252.276   &	 38.271\\
28  &  Hamburg                  & DePre	 &   -0.282	  &   0.244	  &  -0.013	  &   0.373	 &   49.169	 &  267.298   &	116.614\\
29  &  Motopi Pan				& DePre  &   -0.404	  &   0.055	  &  -0.078	  &   0.415	 &   82.425	 &  259.072	  & 251.853\\	
30  &  Ozerki				    & DePre  &   -0.086	  &  -0.151	  &  -0.289	  &   0.337	 &  116.565	 &  196.552	  & 269.400\\	
31  &  Arpu Kuilpu				& AsPre  &   -0.383	  &   0.224	  &   0.043	  &   0.446	 &   59.794	 &  276.379	  & 250.376\\	
32  &  Flensburg				& AsPost &    0.482	  &   0.188	  &   0.141	  &   0.536	 &   69.445	 &   73.659	  & 349.212\\	
33  &  Cavezzo                  & DePost &	  0.006	  &   0.199	  &  -0.084	  &   0.216	 &   22.960	 &  175.900   &	100.523\\
34  &  Novo Mesto               & DePost &	  0.618	  &  -0.056	  &  -0.147	  &   0.638	 &   95.027	 &  103.368   &	158.978\\
35  &  Madura Cave              & DePre	 &   -0.277	  &  -0.108	  &  -0.002	  &   0.297	 &  111.220	 &  269.609   &	268.703\\
36  &  Traspena				    & DePre  &   -0.398	  &  -0.025	  &  -0.079	  &   0.407	 &   93.518	 &  258.751	  & 117.833\\	
37  &  Winchcombe				& AsPost &    0.056	  &   0.267	  &   0.010	  &   0.273	 &   12.045	 &   79.694	  & 160.241\\	
38  &  Antonin				    & DePre  &   -0.212	  &  -0.057	  &  -0.417	  &   0.471	 &   96.894	 &  206.993	  & 292.575\\	

		\hline
	\end{tabular}
\end{table*}

\begin{table*}
	\centering
	\caption{Nominal orbits of the 19 candidate NEAs progenitors at epoch 2460000.5 JD.}
	\label{tab:neo_orbits}
        \setlength\tabcolsep{2pt} % default value: 6pt
	\begin{tabular}{lccccccc} 
		\hline
	Designation & $a$ (au) & $e$ & $i$ ($^\circ$) & $\Omega$ ($^\circ$) & $\omega$ ($^\circ$) & $M$ ($^\circ$)\\
		\hline
2016 RX    & 2.564762534806265   & 0.7430159649322464   & 13.32038927171481   & 163.6869477168528   & 92.94244137657836   & 219.1642830492932 \\
2016 SL2   & 2.062474771961578   & 0.7694417719511251   & 1.81036115435297    & 17.57269810440301   & 90.90339137084725   & 343.6082890477512 \\
2021 FB    & 0.8427246777483847  & 0.2545728694223879   &  9.748690090403196  & 358.4518153292897   & 28.7554149600137    & 315.2583299127406 \\
2022 UF    & 2.5730036883023004  & 0.63341114989874336  & 0.36753632746543397 & 218.37443715719363  & 131.38839048745186  & 37.406284232043475 \\
2021 PZ1   & 2.364634799369776   & 0.6005267765114393   & 0.3837223115640573  & 312.3287880862847   & 43.02588027657282   & 145.221744238352\\
2021 NV5   & 2.4232937397214642  & 0.60395245231432071  & 1.7146217533748498  & 286.87427481414164  & 37.913588036324398  & 147.52756370586465\\
2019 ST2   & 2.360713484135783   & 0.5936746616690849   & 0.9134261986212852  & 148.3443111992481   & 260.5757535991638   & 350.1867624028299\\
2022 RQ    & 1.4635773482255161  & 0.49759911285260583  & 1.4321781313495281  & 281.20578540149700  & 148.86736375858436  & 62.584884321126040 \\
2013 BR15  & 1.554708606396463   & 0.5204835798046532   & 1.95477797282755     & 102.901345532338   &  284.8925546883462   & 32.86431971445231\\
482488    & 2.4599816749788479  & 0.67863969498218224  & 10.164746072164734  & 209.59680478373903  &  62.360038457242275 &  270.76914909176315\\
2014 KF22  & 1.4975257070308479  & 0.42062292021733566  & 4.9474665864368523  &  68.058161464574212 & 237.29322439428816  & 253.92446624875143 \\
2021 JN2   & 1.173687000118325   & 0.3328521597176412   & 2.569641121995291   & 90.88855796610243   & 33.76881352655698   & 210.6776283527358 \\
2017 FZ64  & 1.93653518628377    & 0.4973485243004753   & 7.654708065751632   & 163.3336464620337   & 52.71333823686388   & 60.912511625351 \\
2021 WT    & 2.388580875809223   & 0.6740508905169703   & 3.450236129095224   & 60.66733234698395   & 298.9417278570344   & 132.9039422465572 \\
2005 VE7   & 2.3322473963386656  & 0.71721231566531540  & 7.5150458469052781  & 287.12863098453755  &  14.946913926937720 &  330.01202768332274\\
454100    & 1.3314136872194222  & 0.418262476165029    & 4.543665801377      & 24.781742651553 & 298.166430154473  & 331.192762486796 \\
2017 MC3   & 1.1384900776606286  & 0.655699554855858    & 7.073023632531      &111.187680237372 & 300.270880329344  & 183.467292685508 \\
2009 FZ4   & 1.1218473724744065  & 0.249500520614930    & 4.200036083824      &357.105648161701 & 288.875765616387  & 178.952261478393 \\
2022 QJ2   & 2.6048219825436361  & 0.654500985555764    & 1.610888877536      &289.421112283214 & 342.207267115725  &  54.547932515782 \\
		\hline
	\end{tabular}
\end{table*}

\begin{figure*}
    \centering
    \includegraphics[width=0.48\textwidth]{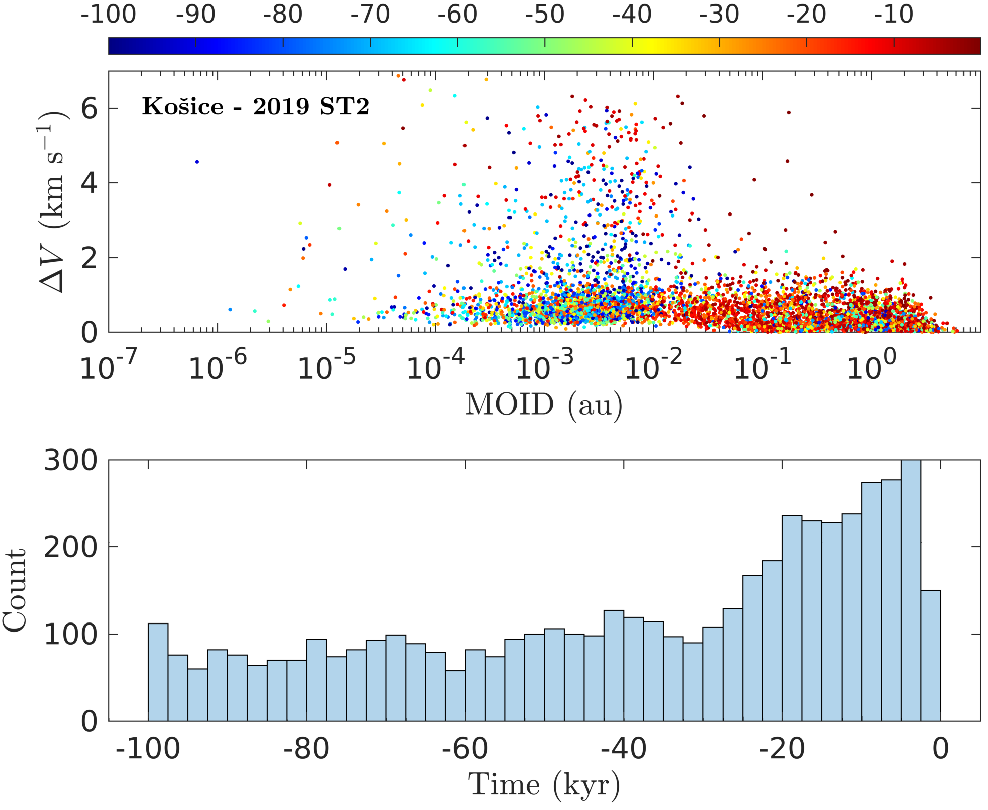}
    \includegraphics[width=0.48\textwidth]{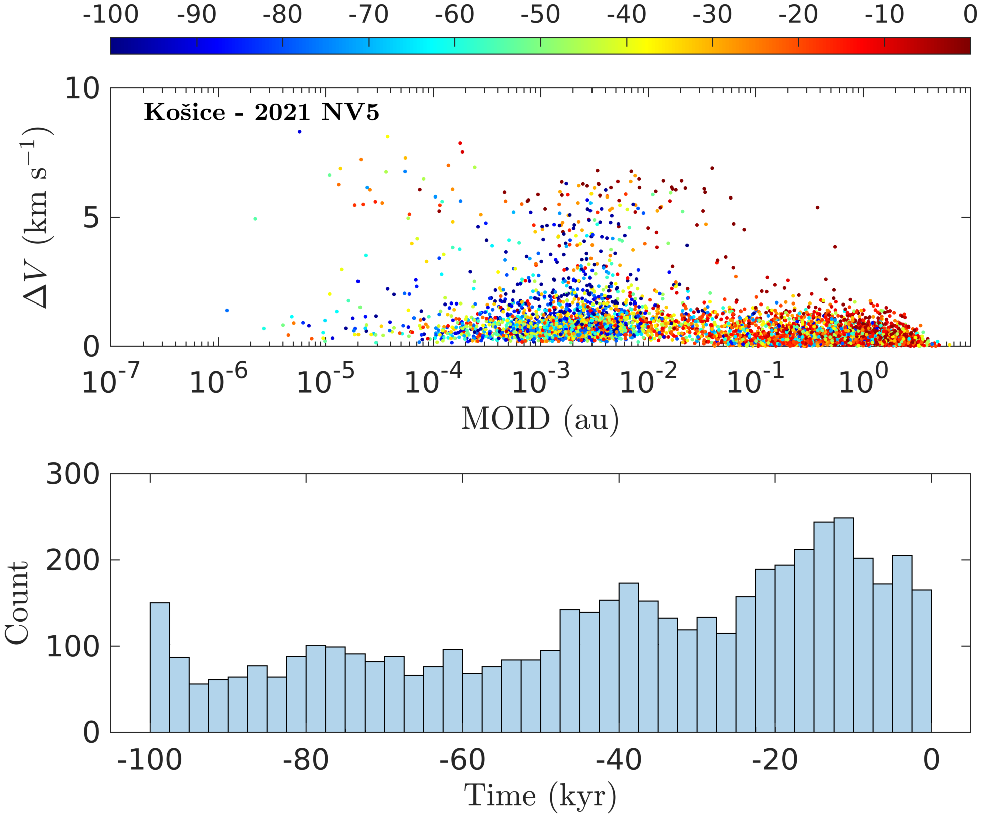}
    \vskip 10pt
    \includegraphics[width=0.48\textwidth]{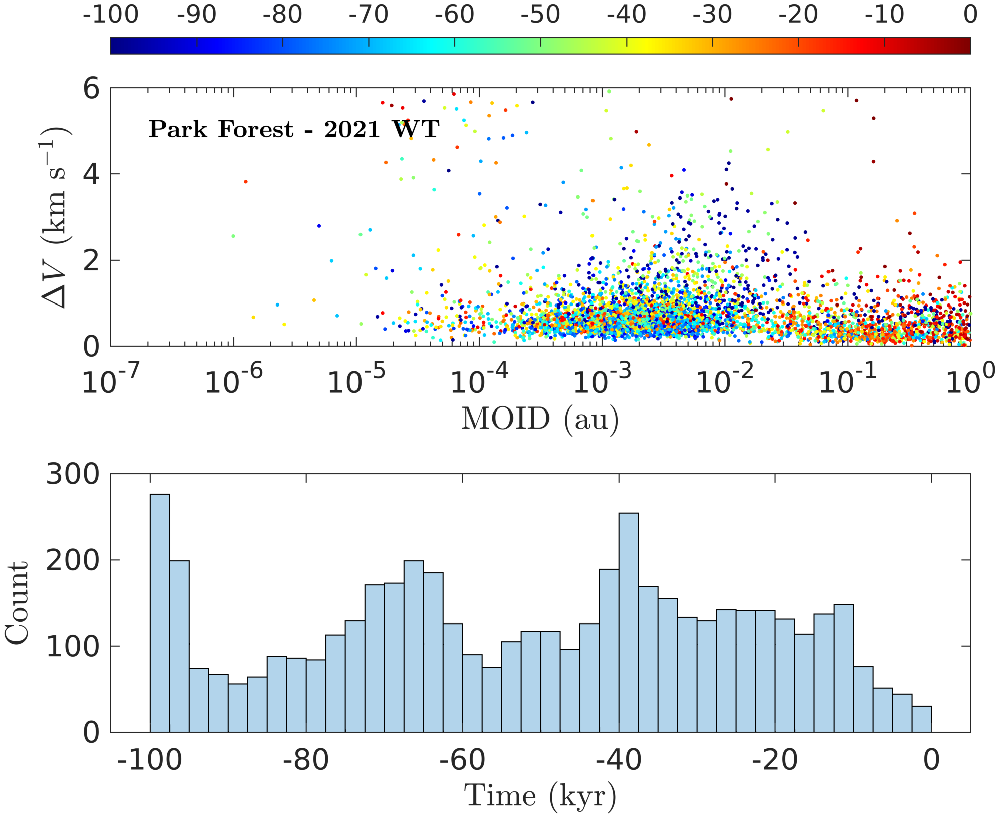}
    \includegraphics[width=0.48\textwidth]{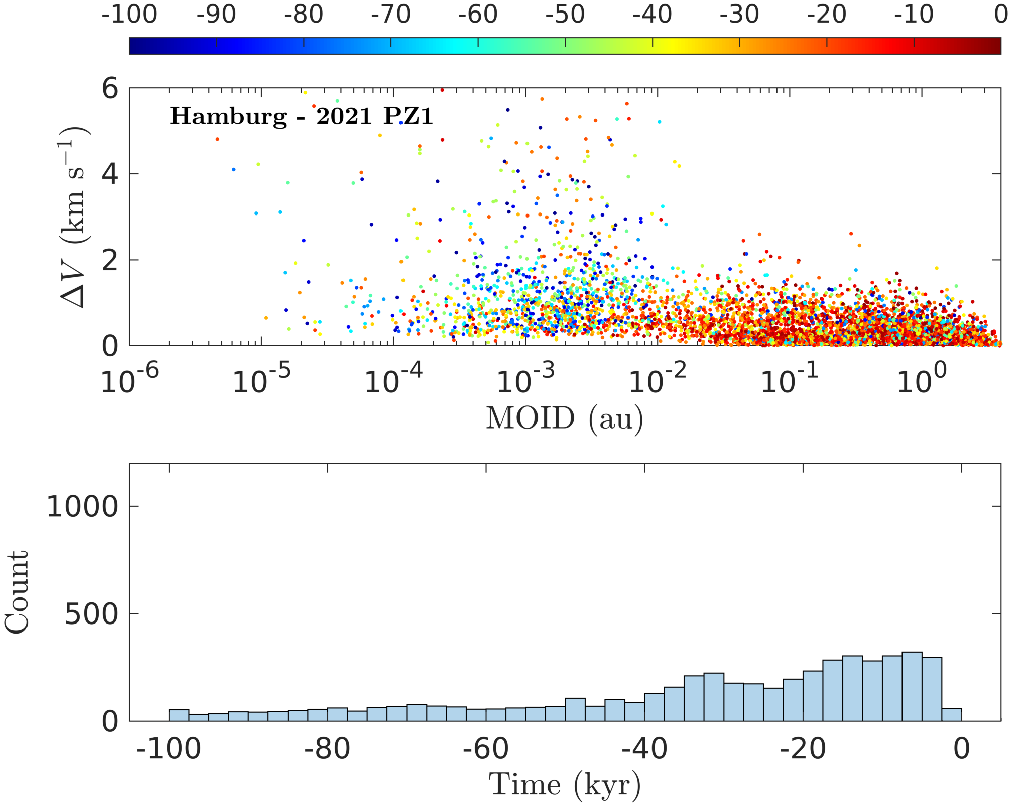}
    \vskip 10pt
    \includegraphics[width=0.48\textwidth]{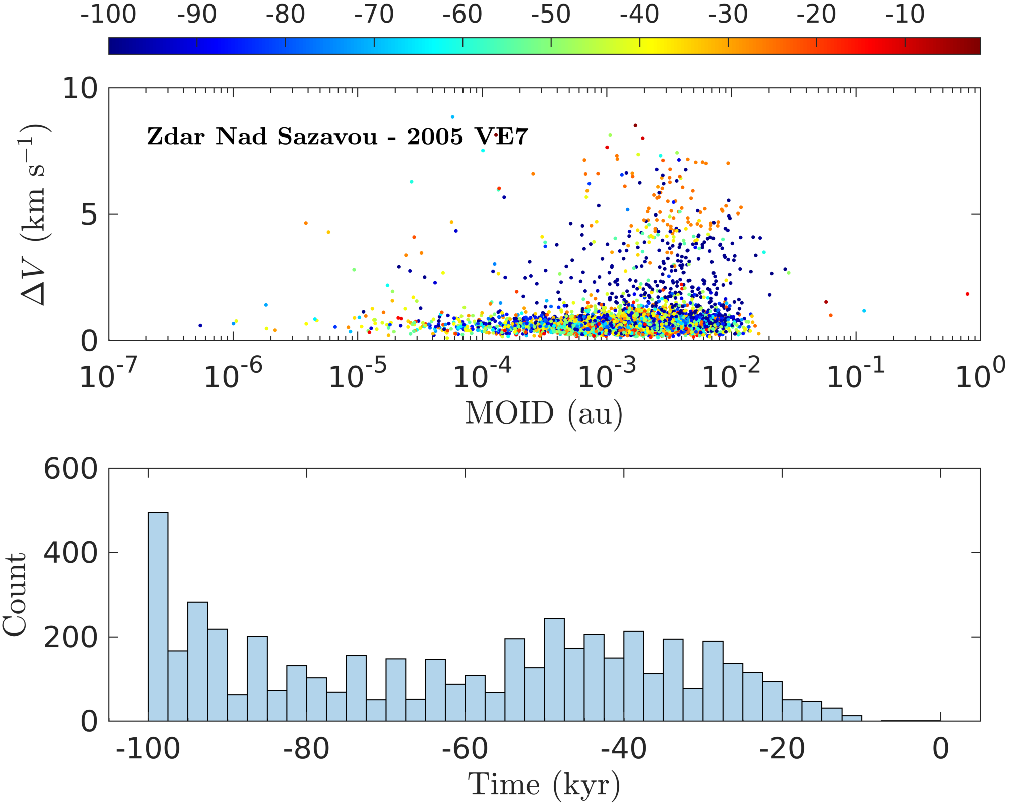}
    \includegraphics[width=0.48\textwidth]{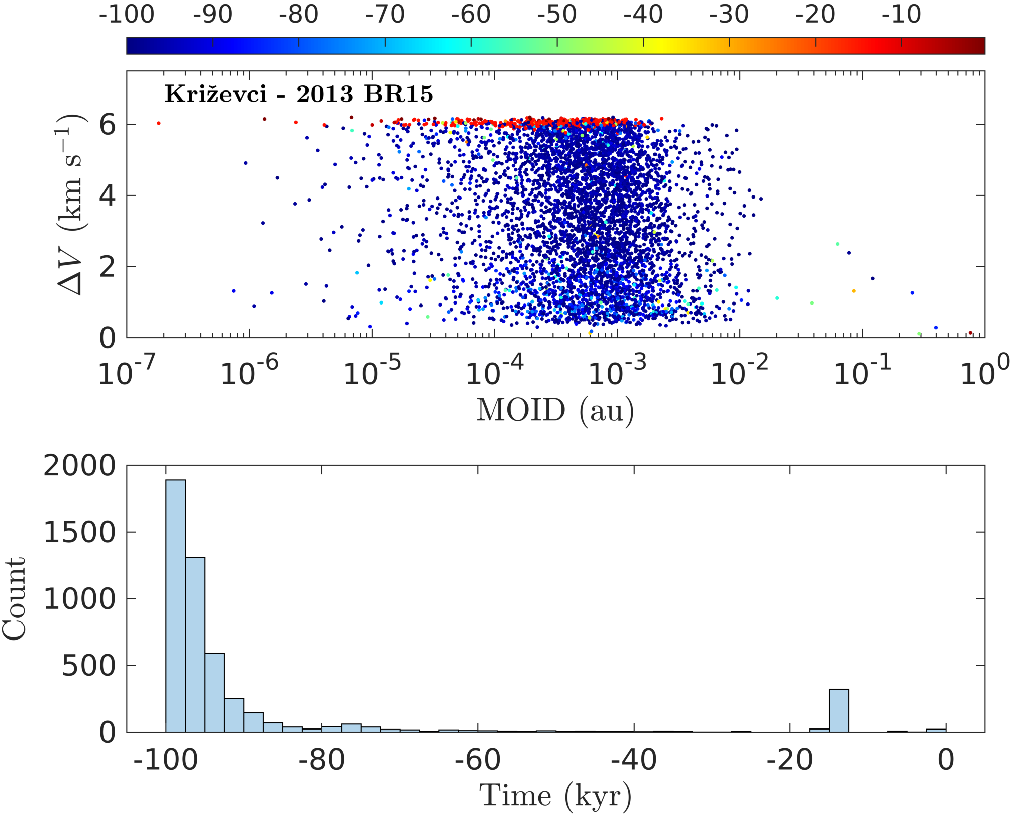}
    \caption{Distributions of the relative velocity vs. the MOID at the time $t_{\min}$ and distribution of $t_{\min}$, for the pairs Košice$-$2021 VN5, Košice$-$2019 ST2, Park Forest$-$2021 WT, Hamburg$-$2021 PZ1, Žd'ár nad Sázavou$-$2005 VE7, and Križevci$-$2013 BR15.}
    \label{fig:Pairs_Appendix}
\end{figure*}

%%%%%%%%%%%%%%%%%%%%%%%%%%%%%%%%%%%%%%%%%%%%%%%%%%

%%%%%%%%%%%%%%%%% APPENDICES %%%%%%%%%%%%%%%%%%%%%

%\appendix


%%%%%%%%%%%%%%%%%%%%%%%%%%%%%%%%%%%%%%%%%%%%%%%%%%

% Don't change these lines
\bsp	% typesetting comment
\label{lastpage}
\end{document}